\documentclass[twocolumn,epjc3]{svjour3}
\usepackage{graphics}

\usepackage{multicol}
\usepackage{cuted}
\usepackage{flushend}

\usepackage{microtype}
\usepackage{bbm}
\usepackage{amsmath}
\usepackage{mathtools}
\usepackage{caption}
\usepackage{subcaption}
\usepackage{amssymb}
\usepackage{mathrsfs}       
\usepackage[pdftex]{graphicx}
\usepackage{pgfplots}
\pgfplotsset{width=10cm,compat=1.9}
\usepackage{xargs}       
\usepackage{wasysym}
\usepackage{engrec}
\usepackage{enumerate}
\usepackage{appendix}
\usepackage{empheq}
\usepackage{float}
\usepackage{stmaryrd}
\usepackage[parfill]{parskip}
\usepackage[pdftex,breaklinks]{hyperref}
\usepackage{titletoc}
\usepackage{makecell}
\usepackage{lscape}
\usepackage{algorithm}
\usepackage{algorithmicx}
\usepackage{tensor}
\usepackage{algpseudocode}
\usepackage{blkarray}
\usepackage{multirow}
\usepackage{bigstrut}
\usetikzlibrary{plotmarks}

\usepackage{simpler-wick}
\usepackage{bm}

\newenvironment{customlegend}[1][]{%
    \begingroup
    \csname pgfplots@init@cleared@structures\endcsname
    \pgfplotsset{#1}%
}{%
    \csname pgfplots@createlegend\endcsname
    \endgroup
}%
\def\addlegendimage{\csname pgfplots@addlegendimage\endcsname}

\definecolor{mygreen}{RGB}{21, 158, 10}
\begin{document}
\title{On the off-diagonal Wick's theorem and  Onishi formula}
\subtitle{Alternative and consistent approach to off-diagonal operator and norm kernels}
\author{A. Porro\thanksref{ad:saclay} \and T. Duguet\thanksref{ad:saclay,ad:kul}}

\institute{
\label{ad:saclay}
IRFU, CEA, Universit\'e Paris-Saclay, 91191 Gif-sur-Yvette, France 
\and
\label{ad:kul}
KU Leuven, Department of Physics and Astronomy, Instituut voor Kern- en Stralingsfysica, 3001 Leuven, Belgium 
}

\date{Received: \today{} / Revised version: date}

\maketitle
%
%
\begin{abstract}
The projected generator coordinate method based on the configuration mixing of non-orthogonal Bogoliubov product states\footnote{The simpler projected generator coordinate method based on non-orthogonal Slater determinants is denoted as the non-orthogonal configuration interaction (NOCI) method in quantum chemistry~\cite{burton09a}.}, along with more advanced methods based on it, require the computation of off-diagonal Hamiltonian and norm kernels. While the Hamiltonian kernel is efficiently computed via the off-diagonal Wick theorem of Balian and Brezin, the norm kernel relies on the Onishi formula (or equivalently the Pfaffian formula by Robledo or the integral formula by Bally and Duguet). Traditionally, the derivation of these two categories of formulae rely on different formal schemes. In the present work, the formulae for the operator and norm kernels are computed consistently from the same diagrammatic method. The approach further offers the possibility to address kernels involving more general states in the future.
\end{abstract}

\section{Introduction}
\label{intro}

The projected generator coordinate method (PGCM) is a popular and versatile many-body method based on the mixing of Bogoliubov vacua typically generated by solving constrained Hartree-Fock-Bogoliubov (HFB) mean-field equations~\cite{RiSc80}. The PGCM is traditionally employed with empirical effective interactions tailored to the full one-body Hilbert space~\cite{bender03b,Niksic:2011sg,Robledo:2018cdj} or to a so-called valence space~\cite{Gao:2015dla,Jiao:2017opc,Bally:2019miu,Shimizu:2021ltl,Sanchez-Fernandez:2021nfg}. However, the PGCM has recently been extented to the context of {\it ab initio} calculations aiming at approximating exact solutions of Schr{\"o}dinger's equation in the low-energy sector of the $A$-body Hilbert space starting from realistic nuclear Hamiltonians rooted into quantum chromodynamics. Possibly combined with a pre-processing of the Hamiltonian via a multi-reference in-medium similarity renormalization group transformation~\cite{Yao:2019rck,Yao:2018qjv,Frosini:2021ddm}, the PGCM is either exploited as a stand-alone method for nuclear spectroscopy~\cite{Frosini:2021sxj} or as a starting point of an expansion method towards the exact solution~\cite{Frosini:2021fjf,Frosini:2021ddm}.

The PGCM is based on the wave-function ansatz
\begin{equation}
    | \Psi_\nu \rangle \equiv \sum_{r} f_\nu(r) | \Phi(r) \rangle\, , \label{GCMansatz}
\end{equation}
where $\{ | \Phi(r) \rangle\}$ denotes a set of non-orthogonal\footnote{Some of the Bogoliubov states mixed in the PGCM ansatz may be either manifestly or accidentally orthogonal. This situation can be dealt with at the price of a generalization of the situation discussed in the present work where any pair of Bogoliubov states entering Eq.~\eqref{GCMansatz} are considered to be non-orthogonal.} Bogoliubov states labelled by the collective coordinate\footnote{The collective coordinate is multi dimensional and contains the variable(s) parameterizing the transformations associated with the symmetry(ies) being restored via projection techniques.} $r$.  The unknown\footnote{The part of the coefficients fixed by the structure of the symmetry group does not have to be determined variationally.} coefficients $\{f_\nu(r)\}$ are determined on the basis of Ritz' variational principle, the energy minimization associated with $| \Psi_\nu \rangle$ leading to the well-known Hill-Wheeler-Griffin (HWG) secular equation
\begin{equation}\label{eq:HWG}
    \sum_r\Big[\langle \Phi(l)|H|\Phi(r) \rangle-E_\nu\langle \Phi(l)|\Phi(r) \rangle \Big]f_\nu(r)=0\,,
\end{equation}
which can be rewritten as
\begin{equation}\label{eq:HWG_ratio}
    \sum_r\Big[\frac{\langle \Phi(l)|H|\Phi(r) \rangle }{\langle \Phi(l)|\Phi(r) \rangle}-E_\nu\Big]\langle \Phi(l)|\Phi(r) \rangle f_\nu(r)=0\,,
\end{equation}
where $H$ is the Hamiltonian and where $E_\nu$ denotes the set of PGCM energies approximating a subset of its eigenvalues.

The key ingredients entering Eq.~\eqref{eq:HWG} and the computation of observables are the so-called off-diagonal connected operator and norm kernels associated with two Bogoliubov states $\langle \Phi(l)|$ and  $|\Phi(r)\rangle$. Given a generic operator $O$, the operator and norm kernels are respectively given by
\begin{subequations}
\label{kernels}
\begin{align}
\mathcal{O}(l,r) &\equiv \langle \Phi(l)|O|\Phi(r) \rangle \, , \label{kernelO} \\
\mathcal{N}(l,r)&\equiv \langle \Phi(l)|\Phi(r) \rangle \,, \label{kernelN}
\end{align}
\end{subequations}
out of which the {\it connected} operator kernel is defined through 
\begin{align}
{o}(l,r) &\equiv  \frac{\mathcal{O}(l,r)}{\mathcal{N}(l,r)} \, . \label{connectedkernel}
\end{align}

The connected operator kernel is efficiently computed via the off-diagonal Wick theorem (ODWT) of Balian and Brezin~\cite{Balin69wick}. The norm kernel relies on the Onishi formula~\cite{onishi66}, on the Pfaffian formula by Robledo~\cite{Robledo:2009yd} or the integral formula by Bally and Duguet~\cite{Bally:2017nom}. Traditionally, the derivation of these two categories of formulae relies on different formal schemes that do not seem to share a common ground. One exception relies on the use of fermion coherent states based on Grassmann variables allowing one to express both the connected operator kernel~\cite{Mizusaki:2012aq,Mizusaki:2013dda} and the norm kernel~\cite{Robledo:2009yd} in terms of Pfaffians. The goal of the present work is to provide another consistent derivation of the connected operator kernel, i.e. of the ODWT, and of the norm kernel based on a common diagrammatic method.  

The paper is organized as follows. Section~\ref{sec_basics} introduces the necessary elements of formalisms.  In Sec.~\ref{comput_kernels}, the diagrammatic method is used to derive both the ODWT for the connected operator kernel and the Onishi/Pfaffian formula for the norm kernel. Conclusions and perspectives are elaborated on in Sec.~\ref{conclusions}. The paper is complemented with several appendices providing necessary technical details, i.e. the normal-ordered representation of operators, the detailed diagrammatic rules and the standard derivation of the ODWT for completeness and comparison.

\section{Basics of Bogoliubov algebra}
\label{sec_basics}

The core of the present work involves three non-orthogonal Bogoliubov states\footnote{While a generalization is possible, the reference state $| \Phi \rangle$ is supposed to be non-orthogonal to both $|\Phi(l)\rangle$ and $|\Phi(r)\rangle$.} denoted as $|\Phi(l)\rangle$,   $|\Phi(r)\rangle$ and  $|\Phi\rangle$. 

\subsection{Bogoliubov vacuum}

Each of these three states is a vacuum for a set of associated quasi-particle operators. Taking $| \Phi \rangle$ as an example, this property reads as
\begin{equation}
    \beta_k | \Phi \rangle = 0 \quad\forall k\, , 
\end{equation}
where the set of quasi-particle operators $\{\beta^{\dagger}_k, \beta_k\}$ is related to particle operators $\{c^{\dagger}_a, c_a\}$ associated with an arbitrary basis of the one-body Hilbert space ${\cal H}_1$ via a unitary Bogoliubov transformation of the form
\begin{subequations}
\label{eq:Bogo_transf_ext}
    \begin{align}
\beta_k&\equiv\sum_a\big(\mathcal{U}^*_{ak}\,c_a + \mathcal{V}^*_{ak}\,c_a^\dagger\big)\,, \\
\beta_k^\dagger&\equiv\sum_a\big(\mathcal{U}_{ak}\,c_a^\dagger + \mathcal{V}_{ak}\,c_a\big)\, .
    \end{align}
\end{subequations}
This transformation can be written more compactly via a matrix representation
\begin{equation}
    \begin{pmatrix}
    \beta\\\beta^\dagger
    \end{pmatrix}=\mathcal{W}^\dagger\begin{pmatrix}
    c\\c^\dagger
    \end{pmatrix}\,,
\end{equation}
where the Bogoliubov matrix
\begin{equation}
    \mathcal{W}\equiv\begin{pmatrix}
    \mathcal{U} & \mathcal{V}^*\\
    \mathcal{V} & \mathcal{U}^*
    \end{pmatrix}
\end{equation}
is unitary, such that the following relations hold
\begin{equation}
    \mathcal{W}\mathcal{W}^\dagger=\mathcal{W}^\dagger\mathcal{W}=1\,.\label{eq:ferm_unitary}
\end{equation}
This condition implies that the canonical fermionic anticommutation rules valid for the particle operators propagate to quasi-particle ones.

\subsection{Thouless theorem}
\label{secthouless}

Starting from Bogoliubov transformations $\mathcal{W}(l)$, $\mathcal{W}(r)$ and $\mathcal{W}$ defining the three sets of quasi-particle operators $\{\beta^{\dagger}_k(l), \beta_k(l)\}$,  $\{\beta^{\dagger}_k(r), \beta_k(r)\}$ and  $\{\beta^{\dagger}_k, \beta_k\}$, respectively, Thouless' theorem~\cite{thouless60} allows one to connect the three vacua $|\Phi(l)\rangle$,   $|\Phi(r)\rangle$ and  $|\Phi\rangle$.

First, $| \Phi(l) \rangle$ and  $| \Phi(r) \rangle$ are expressed with respect to $| \Phi \rangle$. Taking $| \Phi(r) \rangle$ as an example, the transformation connecting the two sets of quasi-particle operators is given by
\begin{align}
    \begin{pmatrix}
    \beta(r)\\
    \beta^\dagger(r)
    \end{pmatrix} 
    &=\mathcal{W}^\dagger(r)\mathcal{W}
    \begin{pmatrix}
    \beta\\\beta^\dagger
    \end{pmatrix} \nonumber \\ 
    &\equiv   
    \begin{pmatrix}
		U^{\dagger}(r)&V^{\dagger}(r)\\
		V^{T}(r)&U^{T}(r)
	\end{pmatrix}
    \begin{pmatrix}
    \beta\\\beta^\dagger
    \end{pmatrix}  \label{transitionBogo1}
    \,,
\end{align}
with 
	\begin{subequations}
		\begin{align}
			U(r)&\equiv\mathcal{V}^T\mathcal{U}(r)+\mathcal{U}^T\mathcal{V}(r)\,,\\
			V(r)&\equiv\mathcal{U}^\dagger\mathcal{U}(r)+\mathcal{V}^\dagger\mathcal{V}(r)\,.
		\end{align}
	\end{subequations}
Introducing the skew-symmetric matrix
\begin{equation}
\textbf{z}(r)\equiv V^{\ast}(r)U^{\ast-1}(r)\,,
\end{equation}
Thouless' theorem allows one to write 
\begin{subequations}
\label{Thouless1}
    \begin{align}
        | \Phi(r) \rangle &= \langle \Phi|\Phi(r) \rangle e^{\textbf{Z}^{20}(r)} | \Phi\rangle \,,\label{eq:Thouless1}
    \end{align}
where the one-body Thouless operator reads as
\begin{align}
    \textbf{Z}^{20}(r)&\equiv\frac{1}{2}\sum_{k_1k_2}\textbf{z}^{k_1k_2}(r)\beta^\dagger_{k_1}\beta^\dagger_{k_2}\,. \label{thoulessoperator1}
\end{align}
\end{subequations}

Similarly, the transformation
\begin{align}
    \begin{pmatrix}
    \beta(r)\\
    \beta^\dagger(r)
    \end{pmatrix} 
    &=W^\dagger(r)W(l)
    \begin{pmatrix}
    \beta(l)\\\beta^\dagger(l)
    \end{pmatrix}  \nonumber\\ 
    &\equiv   
W^\dagger(l,r)    \begin{pmatrix}
    \beta(l)\\\beta^\dagger(l)
    \end{pmatrix}  \nonumber\\
    &\equiv   
    \begin{pmatrix}
		B^{\dagger}(l,r)&A^{\dagger}(l,r)\\
		A^{T}(l,r)&B^{T}(l,r)
	\end{pmatrix}
    \begin{pmatrix}
    \beta(l)\\\beta^\dagger(l)
    \end{pmatrix} \label{transitionBogo3}
    \,,
\end{align}
with 
	\begin{subequations}
		\begin{align}
			A(l,r)&\equiv V^T(l)U(r)+U^T(l)V(r) \, , \\
			B(l,r)&\equiv U^\dagger(l)U(r)+V^\dagger(l)V(r) \, ,
		\end{align}
	\end{subequations}
leads to defining the Thouless matrix
\begin{equation}
\textbf{z}(l,r)\equiv A^{\ast}(l,r)B^{\ast-1}(l,r)\, , \label{thoulessmatrix3}
\end{equation}
thanks to which $| \Phi(r) \rangle$ can be expressed with respect to $| \Phi(l) \rangle$ according to
\begin{subequations}
\label{Thouless3}
    \begin{align}
        | \Phi(r) \rangle&= \langle \Phi(l)|\Phi(r) \rangle e^{\textbf{Z}^{20}(l,r)} | \Phi(l) \rangle\,,\label{eq:Thouless3}
    \end{align}
where
\begin{align}
    \textbf{Z}^{20}(l,r)&\equiv\frac{1}{2}\sum_{k_1k_2}\textbf{z}^{k_1k_2}(l,r)\beta^\dagger_{k_1}(l)\beta^\dagger_{k_2}(l)\,. \label{thoulessoperator3}
\end{align}
\end{subequations}

\subsection{Elementary contractions}

Given  $| \Phi(l) \rangle$ and $| \Phi(r) \rangle$, the four one-body off-diagonal elementary contractions  are defined in the quasi-particle basis of  $| \Phi \rangle$ through
\begin{eqnarray}
{\bf R}_{k_1k_2}(l,r) &\equiv& 
\left(
\begin{array} {cc}
\frac{\langle \Phi(l) | \beta^{\dagger}_{k_2}\beta^{\phantom{\dagger}}_{k_1}\, | \Phi(r) \rangle}{\langle \Phi(l) | \Phi(r) \rangle} & \frac{\langle \Phi(l) | \beta^{\phantom{\dagger}}_{k_2}\beta^{\phantom{\dagger}}_{k_1} | \Phi(r) \rangle}{\langle \Phi(l) | \Phi(r) \rangle} \nonumber \\
\frac{\langle \Phi(l) | \beta^{\dagger}_{k_2}\beta^{\dagger}_{k_1} | \Phi(r) \rangle}{\langle \Phi(l) | \Phi(r) \rangle} &  \frac{\langle \Phi(l) | \beta^{\phantom{\dagger}}_{k_2}\beta^{\dagger}_{k_1} | \Phi(r) \rangle}{\langle \Phi(l) | \Phi(r) \rangle}
\end{array}
\right)  \\
&\equiv& 
\left(
\begin{array} {cc}
+\rho_{k_1k_2}(l,r) & +\kappa_{k_1k_2}(l,r) \\
-\bar{\kappa}^{\ast}_{k_1k_2}(l,r) &  -\sigma^\ast_{k_1k_2}(l,r)
\end{array}
\right)\, , \label{generalizeddensitymatrix2}
\end{eqnarray}
and satisfy, due to anticommutation rules and complex conjugation,
\begin{subequations}
\label{densitymatrices}
\begin{eqnarray}
\rho^*_{k_1k_2}(l,r) &=& +\rho_{k_2k_1}(r,l)  \, ,         \label{densitymatrices1} \\
\kappa_{k_1k_2}(l,r) &=& - \kappa_{k_2k_1}(l,r) \, ,            \label{densitymatrices2} \\
\bar{\kappa}_{k_1k_2}(l,r) &=& +\kappa_{k_1k_2}(r,l) \, ,       \label{densitymatrices4} \\
\sigma_{k_1k_2}(l,r) &=& +\rho_{k_1k_2}(r,l) - \delta_{ab} \, . \label{densitymatrices3} 
\end{eqnarray}
\end{subequations}

Setting $| \Phi(l) \rangle=| \Phi(r) \rangle=| \Phi \rangle$, one can trivially obtain the diagonal contractions associated with $| \Phi \rangle$ as
\begin{eqnarray}
{\bf R}_{k_1k_2}
&=& 
\left(
\begin{array} {cc}
+\rho_{k_1k_2} & +\kappa_{k_1k_2} \\
-\bar{\kappa}^{\ast}_{k_1k_2} &  -\sigma^\ast_{k_1k_2}
\end{array}
\right) \nonumber \\
&=& 
\left(
\begin{array} {cc}
0 & 0 \\
0 &  \delta_{k_1k_2}
\end{array}
\right)\, . \label{generalizeddensitymatrix3}
\end{eqnarray}

From this most simplistic case, one can easily infer the diagonal contractions associated with $| \Phi(l) \rangle$ in the quasi-particle basis of $| \Phi \rangle$, e.g.
\begin{eqnarray}
{\bf R}_{k_1k_2}(l,l)
&=& 
\left(
\begin{array} {cc}
(V^{\ast}(l)V^{T}(l))_{k_1k_2} &  
(V^{\ast}(l)U^{T}(l))_{k_1k_2} \\
(U^{\ast}(l)V^{T}(l))_{k_1k_2} &  
(U^{\ast}(l)U^{T}(l))_{k_1k_2}
\end{array}
\right)\, . \label{generalizeddensitymatrix4} 
\end{eqnarray}

\section{Computation of the kernels} 
\label{comput_kernels}

The traditional way to compute the connected operator kernel, i.e. to derive the off-diagonal Wick theorem~\cite{Balin69wick}, invokes an {\it asymmetric approach} that consists of expressing, e.g., $| \Phi(r) \rangle $ with respect to $| \Phi(l) \rangle $ via Eq.~\eqref{Thouless3}. This delivers the connected operator kernel under the asymmetric form
\begin{align}
o(l,r) &=  \langle \Phi(l)|Oe^{\textbf{Z}^{20}(l,r)}|\Phi(l) \rangle \, . \label{asymmetricconnectedkernel}
\end{align}
The proof of the off-diagonal Wick theorem~\cite{Balin69wick} based on Eq.~\eqref{asymmetricconnectedkernel} is recalled for reference in \ref{standardGWT}. This constitutes the simplest derivation of the ODWT because the power series associated with the exponential appearing on one side of the operator $O$ in Eq.~\eqref{asymmetricconnectedkernel} naturally terminates after a finite number of terms. 

The asymmetric approach cannot provide access to the norm kernel and thus only delivers half of the needed ingredients. Typically, accessing the norm kernel relies on a  {\it symmetric approach}\footnote{An asymmetric approach to the normal kernel based on finding a {\it unitary} transformation between both states exists~\cite{Bally:2017nom}. This however differs from the present discussion based on non-unitary Thouless transformations.} where $| \Phi(l) \rangle $ and $| \Phi(r) \rangle$ are both expressed with respect to a common reference state $| \Phi \rangle$ according to Eq.~\eqref{Thouless1}. 

In this context, it is of interest to consistently derive the connected operator kernel (i.e. the off-diagonal Wick theorem) and the norm kernel (i.e. the Onishi or Pfaffian formula) via a symmetric approach. While this was achieved starting from fermion coherent states based on Grassmann variables~\cite{Mizusaki:2012aq,Mizusaki:2013dda}, this is presently realized using standard diagrammatic techniques. The symmetric approach leads to expressing the operator and norm kernels as
\begin{subequations}
\label{symmetrickernels}
\begin{align}
\frac{\mathcal{O}(l,r)}{\langle \Phi(l)|\Phi \rangle \langle \Phi|\Phi(r) \rangle} &\equiv  \langle \Phi | e^{\textbf{Z}^{20}(l)^\dagger} O e^{\textbf{Z}^{20}(r)} |\Phi \rangle \, , \label{symmetrickernelO} \\
\frac{\mathcal{N}(l,r)}{\langle \Phi(l)|\Phi \rangle \langle \Phi|\Phi(r) \rangle} &\equiv  \langle \Phi | e^{\textbf{Z}^{20}(l)^\dagger} e^{\textbf{Z}^{20}(r)} |\Phi \rangle \,, \label{symmetrickernelN}
\end{align}
\end{subequations}
such that the connected operator kernel itself reads as
\begin{align}
o(l,r) &\equiv  \frac{\langle \Phi | e^{\textbf{Z}^{20}(l)^\dagger} O e^{\textbf{Z}^{20}(r)} |\Phi \rangle}{\langle \Phi | e^{\textbf{Z}^{20}(l)^\dagger} e^{\textbf{Z}^{20}(r)} |\Phi \rangle} \, . \label{symmetricconnectedkernel}
\end{align}
It is worth noting that the above expression resembles the expectation value at play in variational coupled cluster (vCC) theory~\cite{bartlett88a,robinson12a,marie21a}. On the one hand, it is more general because the two involved Thouless operators are not equal. On the other hand, it is more restricted given that the Thouless operators are one-body excitation operators, which essentially corresponds to vCC with singles (vCCS) approximation. Still, it is well known from vCC that, (i) while the norm overlap in the denominator can be exactly cancelled out in the numerator, (ii) the expansions of the two exponentials do not terminate and thus produce an infinite number of terms. It will thus have to be shown how this apparent difficulty can be resolved to compute $\mathcal{N}(l,r)$ and $o(l,r)$ exactly.

\subsection{Operator kernel}

Employing the simplified notations
\begin{subequations}
\begin{align}
    \textbf{R} &\equiv \textbf{Z}^{20}(r) =\frac{1}{2}\sum_{k_1k_2}\textbf{z}^{k_1k_2}(r)\beta^\dagger_{k_1}\beta^\dagger_{k_2}\,, \label{defR} \\
    \textbf{L}&\equiv \textbf{Z}^{20}(l)^{\dagger} = \frac{1}{2}\sum_{k_1k_2}\textbf{z}^{\ast}_{k_1k_2}(l)\beta_{k_2}\beta_{k_1}\,, \label{defL}
\end{align}
\end{subequations}
for the Thouless operators fulfilling $\langle \Phi | \textbf{R} = \textbf{L} | \Phi \rangle = 0$, the operator kernel is expressed as
\begin{align}
\frac{\mathcal{O}(l,r)}{\langle \Phi(l) |\Phi \rangle \langle \Phi| \Phi(r) \rangle} &\equiv \langle \Phi|e^{\textbf{L}} O e^{\textbf{R}} |\Phi \rangle  \nonumber \\
&= \sum_{n=0}^N\sum_{\substack{i,j=0\\i+j=2n}}^{2n} \sum_{s,t=0}^\infty\frac{1}{s!t!}\langle \Phi|\textbf{L}^s \textbf{O}^{ij} \textbf{R}^t|\Phi \rangle \nonumber \\
&\equiv \sum_{n=0}^N\sum_{\substack{i,j=0\\i+j=2n}}^{2n}  \sum_{s,t=0}^\infty \langle _s|^i \, _j| ^t \rangle \, ,  \label{SimplifiedsymmetrickernelO}  \
\end{align}
where $O$ has been decomposed into normal-ordered contributions $\{\textbf{O}^{ij}\}$ with respect to $| \Phi \rangle$; see \ref{NO_operators} for details\footnote{Given the chosen Bogoliubov reference state $| \Phi \rangle$, it is natural to normal order the operator $O$ with respect to that state as is presently done. While this is already very general, one can easily go one step further and express the operator in normal order with respect to yet another product state, e.g. the particle vacuum. The connection between both situations is straightforward.}.

The operator kernel thus takes the form of an infinite sum of expectation values in the Bogoliubov vacuum $| \Phi \rangle$ such that standard Wick's theorem with respect to it straightforwardly applies. As a result, diagrammatic rules can be worked out to efficiently compute the complete set of contributions to each matrix element $\langle _s|^i \, _j| ^t \rangle$. The corresponding diagrammatic is introduced in details in \ref{diagrules}. 

\begin{figure*}[t!]
\centering
   \includegraphics[width=.9\textwidth]{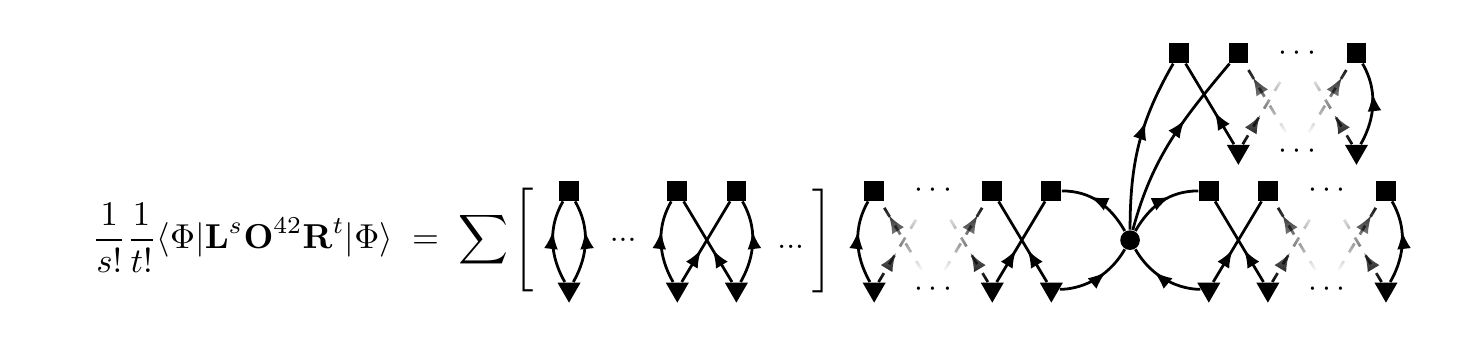}
\caption{
\label{diagramopkernel}
Diagrammatic representation of the matrix element $\langle _s|^4 \, _2| ^t \rangle$, where $s$ and $t$ are positive integers, contributing to the operator kernel $\langle \Phi(l) |\textbf{O}^{42}| \Phi(r) \rangle$. Each square (triangle) vertex represents an operator $\textbf{L}$ ($\textbf{R}$), whereas the dot vertex  denotes the operator $\textbf{O}^{42}$. Whereas the vertex of $\textbf{L}$ ($\textbf{R}$) displays two lines entering (leaving) it, two lines enter the vertex representing $\textbf{O}^{42}$ and four lines leave it. See~\ref{diagrules} for relevant details regarding the diagrammatic representation.}
\end{figure*}

The operators $\textbf{L}$, $\textbf{R}$ and $\textbf{O}^{ij}$ being expressed in the quasi-particle basis associated with $| \Phi \rangle$, the elementary contractions at play in the present application of Wick's theorem take the simple form given by Eq.~\eqref{generalizeddensitymatrix3}. The operator $\textbf{L}$ ($\textbf{R}$) being a pure deexcitation (excitation) operator, no contraction may occur within itself or among its various occurrences. It makes mandatory each $\textbf{L}$ ($\textbf{R}$) operator to contract with either the $i$ creation ($j$ annihilation) operators inside $\textbf{O}^{ij}$ or with the  $\textbf{R}$ ($\textbf{L}$) operators. 

\subsubsection{Factorization of the norm kernel}

Focusing on a single normal-ordered component $\textbf{O}^{ij}$, diagrams making up the operator kernel display characteristic topologies. Indeed, each contribution generated via the application of Wick's theorem is the product of {\it closed strings of contractions}, each of which involves a subset of the $\textbf{L}$ and $\textbf{R}$ operators at play. Whenever $i+j\geq 4$, several strings involve quasi-particle operators belonging to $\textbf{O}^{ij}$, thus forming a super string {\it connected} to the operator $\textbf{O}^{ij}$. Translated into diagrammatic language, each contribution to the operator kernel is thus made out of disjoint closed sub-diagrams, one of which is connected to $\textbf{O}^{ij}$. This key feature is illustrated in Fig.~\ref{diagramopkernel} where the diagrammatic expansion of the matrix element $\langle _s|^4 \, _2| ^t \rangle$ contributing to the operator kernel $\langle \Phi(l) |\textbf{O}^{42}| \Phi(r) \rangle$ is schematically displayed.

Thus, the algebraic contributions to $\langle _s|^i \, _j| ^t \rangle$ can be separated into two distinct factors associated with the two categories of strings. For a given $s$ ($t$) value, $s'$ ($t'$) operators $\textbf{L}$ ($\textbf{R}$) belong to closed strings that are not connected, with $s'$ ($t'$) running from $0$ to $s$ ($t$). For a given $s'$ ($t'$) value, there are $\binom{s}{s'}$ ($\binom{t}{t'}$) ways to select the $\textbf{L}$'s ($\textbf{R}$'s). 
\begin{strip}
Eventually, the matrix elements entering the operator kernel can be written as
\begin{align}
\label{eq:factor_out1}
    \langle \Phi|e^\textbf{L}\textbf{O}^{ij}e^\textbf{R}|\Phi \rangle &=\sum_{s,t=0}^\infty\frac{1}{s!t!}\langle \Phi|\underbrace{\textbf{L}\cdot \textbf{L}\cdot...\cdot \textbf{L}\cdot \textbf{L}}_{\text{$s$ times}}\textbf{O}^{ij}\underbrace{\textbf{R}\cdot \textbf{R}\cdot...\cdot \textbf{R}\cdot \textbf{R}}_{\text{$t$ times}}|\Phi \rangle\nonumber\\
    &=\sum_{s,t=0}^\infty\frac{1}{s!t!}\sum_{s'=0}^s\sum_{t'=0}^t\binom{s}{s'}\binom{t}{t'}\langle\Phi|\underbrace{\textbf{L}\cdot ...\cdot \textbf{L}}_{\text{$(s-s')$ times}}\textbf{O}^{ij}\underbrace{\textbf{R}\cdot ...\cdot \textbf{R}}_{\text{$(t-t')$ times}}|\Phi\rangle_c \langle \Phi|\underbrace{\textbf{L}\cdot ...\cdot \textbf{L}}_{\text{$s'$ times}}\underbrace{\textbf{R}\cdot ...\cdot \textbf{R}}_{\text{$t'$ times}}|\Phi \rangle\,,
\end{align}
where the index $c$ stands for \textit{connected terms}. Reshuffling the sums allows one to factorize in front of each connected contribution the infinite set of disjoint closed contributions making up the norm kernel according to
\begin{align}\label{eq:factor_out2}
    \langle \Phi|e^\textbf{L}\textbf{O}^{ij}e^\textbf{R}|\Phi \rangle &=\sum_{s',t'=0}^\infty\sum_{s=s'}^\infty\sum_{t=t'}^\infty\frac{1}{s'!(s-s')!}\frac{1}{t'!(t-t')!}\langle \Phi|\underbrace{\textbf{L}\cdot ...\cdot \textbf{L}}_{\text{$(s-s')$ times}}\textbf{O}^{ij}\underbrace{\textbf{R}\cdot ...\cdot \textbf{R}}_{\text{$(t-t')$ times}}|\Phi \rangle_c \langle\Phi|\underbrace{\textbf{L}\cdot ...\cdot \textbf{L}}_{\text{$s'$ times}}\underbrace{\textbf{R}\cdot ...\cdot \textbf{R}}_{\text{$t'$ times}}|\Phi \rangle \nonumber\\
    &=\sum_{s,t=0}^\infty\frac{1}{s!t!}\langle \Phi|\underbrace{\textbf{L}\cdot ...\cdot \textbf{L}}_{\text{$s$ times}}\textbf{O}^{ij}\underbrace{\textbf{R}\cdot ...\cdot \textbf{R}}_{\text{$t$ times}}|\Phi\rangle_c\sum_{s',t'=0}^\infty\frac{1}{s'!t'!} \langle \Phi|\underbrace{\textbf{L}\cdot ...\cdot \textbf{L}}_{\text{$s'$ times}}\underbrace{\textbf{R}\cdot ...\cdot \textbf{R}}_{\text{$t'$ times}}|\Phi \rangle \nonumber\\
    &=\langle \Phi|e^\textbf{L} \textbf{O}^{ij}  e^\textbf{R}|\Phi \rangle _c\,\langle \Phi|e^\textbf{L}e^\textbf{R}|\Phi \rangle \, .
\end{align}
\end{strip}%
The above equation demonstrates that the norm kernel exactly factorizes in the operator kernel\footnote{Nothing in the proof depends on the character, e.g. rank, of the operators $\textbf{R}$ and $\textbf{L}$. Thus, the exact factorization of the norm kernel out of the operator kernel constitutes a general result going beyond the scope of the present study that constraints $\textbf{R}$ ($\textbf{L}$) to be a one-body excitation (de-excitation) operator.} such that the connected operator kernel is, hence the name, the sum of connected, necessarily joint, contributions
\begin{align}
\label{eq:connectedkernel_diag}
o(l,r) &= \langle \Phi|e^\textbf{L} O e^\textbf{R}|\Phi \rangle_c = \sum_{n=0}^N\sum_{\substack{i,j=0\\i+j=2n}}^{2n}  \sum_{s,t=0}^\infty \langle _s|^i \, _j| ^t \rangle_c \,,
\end{align}
where, for any $\textbf{O}^{ij}$, the condition
\begin{equation}\label{eq:ij_condition}
    s-t=\frac{i-j}{2}
\end{equation}
is satisfied for each connected matrix element $\langle _s|^i \, _j| ^t \rangle$ given that $\textbf{L}$ ($\textbf{R}$) contains two quasi-particle annihilation (creation) operators. 

In spite of the factorization of the norm kernel, the symmetric approach does not lead to a natural termination of the infinite number of terms making up the connected operator kernel\footnote{The asymmetric approach to the connected operator kernel detailed in \ref{standardGWT}, including the natural termination of the exponential at play, can be recovered from the results obtained below by setting $\textbf{L}\equiv 0$ a posteriori.}. The operators $\textbf{L}$ and $\textbf{R}$ being presently of one-body character, the infinite series thus generated can now be shown to be factorizable in terms of off-diagonal elementary contractions such that the ODWT is recovered.

\begin{strip}
\subsubsection{Off-diagonal Wick's theorem}

The connected operator kernel associated with the normal-order component $\textbf{O}^{ij}$ of arbitrary rank $n\equiv (i+j)/2$ reads as
\begin{align}
\langle \Phi|e^\textbf{L} \textbf{O}^{ij} e^\textbf{R}|\Phi \rangle_c &= \sum_{s,t=0}^\infty \langle _s|^i \, _j| ^t \rangle_c \nonumber \\
&=\sum_{s,t=0}^\infty  \frac{1}{s!t!}
\frac{1}{i!j!} \sum_{\substack{k_1...k_i\\l_1...l_j}} \textbf{o}^{k_1...k_i}_{l_1...l_j} \langle \Phi|\underbrace{\textbf{L}\cdot ...\cdot \textbf{L}}_{\text{$s$ times}}\beta^\dagger_{k_1}...\beta^\dagger_{k_i}\beta_{l_j}...\beta_{l_1}\underbrace{\textbf{R}\cdot ...\cdot \textbf{R}}_{\text{$t$ times}}|\Phi\rangle_c\, , \label{eq:gen_Oij_before}
\end{align}
\end{strip}
where Eq.~\eqref{componentNO_op} has been used.

The calculation of the connected operator kernel relies on the following considerations that are consistent with the diagrammatic rules detailed in \ref{diagrules}
\begin{enumerate}
\item Each contribution to $\langle _s|^i \, _j| ^t \rangle_c$ is made out of strings of contractions connected to quasi-particle operators belonging to $\textbf{O}^{ij}$. The characteristics of the operators $\textbf{L}$ and $\textbf{R}$ strongly constrains the topology of these connected strings
\begin{enumerate}
\item Starting from a quasi-particle operator belonging to $\textbf{O}^{ij}$, a connected string of contractions goes through a set of $\textbf{L}$ and $\textbf{R}$ operators until it reaches other quasi-particle operators of $\textbf{O}^{ij}$.
\item Two successive contractions involving an operator $\textbf{L}$ ($\textbf{R}$) exhaust the two quasi-particle operators it contains. Consequently, a string necessarily forms a single loop connecting two\footnote{Would $\textbf{L}$ and $\textbf{R}$ be of higher rank, e.g. two-body operators, this property would be lost. Indeed, an operator, e.g., $\textbf{L}$ belonging to a loop going through an alternate succession of $\textbf{L}$ and $\textbf{R}$ operators connecting two quasi-particle operators of $\textbf{O}^{ij}$ could further entertain a contraction with an operator $\textbf{R}$ belonging to another closed loop, thus forming a more elaborate closed string eventually involving more than two quasi-particle operators of $\textbf{O}^{ij}$.} quasi-particle operators of $\textbf{O}^{ij}$. 
\item There exist four types of closed strings joining two quasi-particle operators of $\textbf{O}^{ij}$. A string of contractions starting from an operator $\beta^{\dagger}$ and ending at an operator $\beta$ is schematically  indicated below by $[\beta^\dagger\beta]$ whenever the former operator is located to the left of the latter\footnote{Given that $\textbf{O}^{ij}$ is in normal-ordered form, no $[\beta\beta^\dagger]$ string may occur. This would however be the case if the present discussion were extended to the computation of the connected kernel associated with any {\it product}, e.g. $\textbf{O}^{ij}\textbf{O}^{kl}$, of normal-ordered operators. This happens for example when considering kernels involving elementary excitations of $\langle \Phi(l) | $ and/or $| \Phi(r) \rangle$  as in the multi-reference perturbation theory based on a PGCM unperturbed state~\cite{Frosini:2021fjf}. Such an extension is straightforward.}. Similarly, an anomalous string starting from an operator $\beta^{(\dagger)}$ and ending at another operator $\beta^{(\dagger)}$  is denoted as $[\beta\beta]$ ($[\beta^\dagger\beta^\dagger]$). 
\end{enumerate}
\begin{figure}
    \centering
   \includegraphics[width=.35\textwidth]{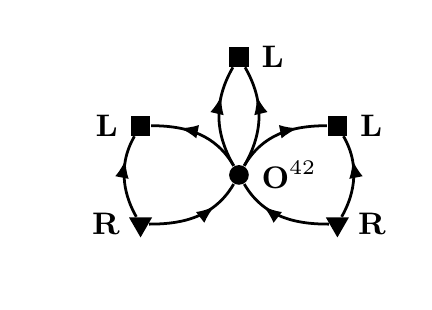}
\caption{
\label{diagramconnectedopkernel}
Example of diagram contributing to the connected operator matrix element $\langle _3|^4 \, _2| ^2 \rangle_c$. In this diagram, one string contains a single operator $\textbf{L}$ connecting two quasi-particle creation operators of $\textbf{O}^{42}$. The two other strings involve one operator $\textbf{L}$ and one operator $\textbf{R}$ each and connect one quasi-particle creation  of $\textbf{O}^{42}$ to one of its quasi-particle annihilation operator.}
\end{figure}
\item The complete set of connected terms contributing to $\langle _s|^i \, _j| ^t \rangle_c$ includes all possible combinations of $n$ normal and anomalous connected loops.  This topological characteristic is responsible for the validity of the off-diagonal Wick's theorem proven below, i.e. for the fact that the end result can be expressed in terms of products of off-diagonal elementary contractions\footnote{If $\textbf{L}$ and/or $\textbf{R}$ are of higher rank, i.e. if $| \Phi(l) \rangle$ and/or $| \Phi(r) \rangle$ do not belong to the manifold of Bogoliubov states, the validity of the off-diagonal Wick's theorem is lost.}. Figure~\ref{diagramconnectedopkernel} displays one such contribution to $\langle _3|^4 \, _2| ^2 \rangle_c$.
\item The combinatorial associated with each contribution to $\langle _s|^i \, _j| ^t \rangle_c$  is obtained from the following considerations
\begin{enumerate}
    \item Among the $n$ strings, $k\leq \text{min}(i,j)$ normal $[\beta^\dagger\beta]$ strings are formed, knowing that $k$ is even whenever $i$ and $j$ are even\footnote{The integers $i$ and $j$ always carry the same parity.} and odd otherwise\footnote{If this rule is not fulfilled, the operator cannot be fully contracted, thus providing a vanishing expectation value by virtue of its normal-ordered form.}. Once $k$ normal strings are formed, there remains an even number $i-k$  $(j-k)$ of quasi-particle creation (annihilation) operators in $\textbf{O}^{ij}$ giving rise to $(i-k)/2$ ($(j-k)/2$) anomalous $[\beta^\dagger\beta^\dagger]$ ($[\beta\beta]$) strings. 
    \item There are $\binom{i}{k}$ different ways to pick $k$ operators out of the $i$ creation operators and, similarly, $\binom{j}{k}$ different ways to pick $k$ operators out of the $j$ annihilation operators. Once this done, there are $k!$ different ways to associate the $k$ creators to the $k$ annihilators to form the $k$ normal strings. Next, there are $(i-k-1)!!$ ($(j-k-1)!!$) possible ways to form the $(i-k)/2$ ($(j-k)/2$) anomalous $[\beta^\dagger\beta^\dagger]$ ($[\beta\beta]$) strings.
    \item For a given $k$, all terms associated with the above combinatorial contribute identically\footnote{Instead of considering all possible strings, this is best seen by keeping the contraction pattern fixed and by exchanging the position of the quasi-particle operators within $\textbf{O}^{ij}$ in all ways consistent with that contraction pattern and by employing the anti-symmetry of the operator matrix elements to recover the original algebraic contribution.} due to the anti-symmetric character of the operator matrix elements under the exchange of any pair of quasi-particle creation (i.e. upper) or annihilation (i.e. lower) indices; see Eq.~\eqref{antisymME}. 
    \item Combining the factor $(i!j!)^{-1}$ originating from the operator with the above combinatorial\footnote{Useful properties of the double factorial are
\begin{subequations}
\begin{alignat}{2}
    n!&=n!!(n-1)!!\quad&&\forall n\\
    n!!&=2^kk!\quad&&\text{for }n=2k\,.
\end{alignat}
\end{subequations}}
one obtains the overall factor
\begin{align}
c(i,j,k) &\equiv \frac{1}{k!}\frac{1}{(i-k)!}\frac{1}{(j-k)!}(i-k-1)!!(j-k-1)!!\, \nonumber
\end{align}
for the diagram associated to $k$ normal strings $[\beta^\dagger\beta]$, $(i-k)/2$ anomalous $[\beta^\dagger\beta^\dagger]$ and  $(j-k)/2$  anomalous strings $[\beta\beta]$. 
\end{enumerate}
\item So far, the focus has been on the nature and the number of closed strings that can be formed in Eq. \eqref{eq:gen_Oij_before} for a given operator $\textbf{O}^{ij}$, while leaving the expansion of $\exp(\textbf{L})$ ($\exp(\textbf{R})$) in the abstract. Let us now consider the specific term of that expansion associated with the power $s$ ($t$), i.e. $\langle _s|^i \, _j| ^t \rangle_c$, and focus on the set of contributions characterized by $k\leq \text{min}(i,j)$ normal strings.
\begin{enumerate}
\item Concentrating first on the $k$ normal strings $[\beta^\dagger\beta]^{(h)}$, a fictitious index $h\in[1,k]$ is introduced to label each of them. There are $s^{(1)}$ operators $\textbf{L}$ out of $s$ involved in the string $[\beta^\dagger\beta]^{(1)}$ and $\binom{s}{s^{(1)}}$ equivalent ways to choose them. Then, there are $s^{(2)}$ such operators out of $(s-s^{(1)})$ involved in $[\beta^\dagger\beta]^{(2)}$ and so on, up to selecting $s^{(k)}$ out of $(s-\sum_{h=1}^{k-1}s^{(h)})$ operators $\textbf{L}$ involved in $[\beta^\dagger\beta]^{(k)}$, such that $s'\equiv\sum_{h=1}^{k}s^{(h)}$ operators out of $s$ are eventually involved in the $k$ normal strings. The overall associated combinatorial factor is given by
\begin{align}
    &\binom{s}{s^{(1)}}\binom{s\text{-}s^{(1)}}{s^{(2)}}\binom{s\text{-}s^{(1)}\text{-}s^{(2)}}{s^{(3)}}\cdots\binom{s\text{-}s^{(1)}\text{-}s^{(2)}\text{-}...\text{-}s^{(k-1)}}{s^{(k)}}\nonumber\\
    &=\frac{s!}{s^{(1)}!s^{(2)}!\ldots s^{(k)}!(s\text{-}s^{(1)}\text{-}s^{(2)}\ldots\text{-}s^{(k)})!}\nonumber\\
    &=\frac{s!}{s^{(1)}!s^{(2)}!\ldots s^{(k)}!(s\text{-}s')!} \, ,
    \label{combinat1}
\end{align}
such that the initial $(s!)^{-1}$ factor is replaced by similar factors for each normal contraction and for the remaining $s-s'$ operators $\textbf{L}$ involved in anomalous strings. The very same operation is considered for the $t'$ operators $\textbf{R}$ involved in the $k$ normal strings, delivering the same combinatorial factor with $s$ variables replaced by $t$ ones. Because one is presently dealing with normal strings, the additional factor $\prod_{h=1}^{k}\delta_{s^{(h)}t^{(h)}}$ must be included given that as many $\textbf{L}$ and $\textbf{R}$ operators must be involved in each of them. 
\item The same reasoning applies to both sets of anomalous strings, knowing that the operators $\textbf{L}$ and $\textbf{R}$ must be selected among the operators that have not been used yet and that the condition $\delta_{s^{(h)}-1,t^{(h)}}$ ($\delta_{s^{(h)},t^{(h)}-1}$) must be used for each string $[\beta^\dagger\beta^\dagger]^{(h)}$  ($[\beta\beta]^{(h)}$). 
\end{enumerate}
\item Eventually summing over all possible values of $s$ and $t$, along with the subset of $s^{(h)}$ and $t^{(h)}$ values, taking into account the Kronecker deltas generated through the processes described above, each contribution to the operator kernel containing  $k\leq \text{min}(i,j)$ normal strings involves intricate sums with the generic structure
    \begin{equation}
        \sum_{a=0}^\infty\sum_{b=0}^{a}\sum_{c=0}^{a-b}\sum_{d=0}^{a-b-c}\ldots \,
    \end{equation}
    where, e.g. for normal strings, $a\equiv n = s = t$, $b\equiv s^{(1)}= t^{(1)}$, $c\equiv s^{(2)}= t^{(2)}$, etc. These intricate sums are easily shown to be equivalent to 
    \begin{equation}
        \sum_{a'=0}^\infty\sum_{b=0}^{\infty}\sum_{c=0}^{\infty}\sum_{d=0}^{\infty}\ldots \, ,
    \end{equation}
    with $a'=a-b-c-d-\ldots$. Each disentangled sum gathers an infinite set of contributions corresponding to closed strings connected to the same pair of quasi-particle operators and involving an increasing number of successive $\textbf{L}$ and $\textbf{R}$ operators. Weighted by the prefactor $(s^{(h)}!)^{-1}(t^{(h)}!)^{-1}$ originating from the successive appropriate applications of Eq.~\eqref{combinat1}, this set of terms exactly make up the corresponding off-diagonal elementary contractions (Eq.~\eqref{generalizeddensitymatrix2}) according to the expansions\footnote{The fourth contraction does not presently occur due to the normal-ordered character of $\textbf{O}^{ij}$.}
\begin{subequations}
\label{expansion_elementarycontractions}
\begin{align}
\rho_{k_1k_2}(l,r) &= \sum_{s,t=0}^\infty\frac{1}{s!t!}\langle \Phi|\textbf{L}^{s} \beta^\dagger_{k_2}\beta_{k_1} \textbf{R}^{t}|\Phi\rangle_c \nonumber \\
& \equiv \sum_{s,t=0}^\infty \langle _s | ^{k_2} | \, |_{k_1} |^t  \rangle_c  \,, \label{expansion_elementarycontractions1} \\
\kappa_{k_1k_2}(l,r) &=  \sum_{s,t=0}^\infty\frac{1}{s!t!}\langle \Phi|\textbf{L}^{s} \beta_{k_2}\beta_{k_1} \textbf{R}^{t}|\Phi\rangle_c \nonumber \\
& \equiv  \sum_{s,t=0}^\infty \langle _s | _{k_2} | \, |_{k_1} |^t  \rangle_c \, \label{expansion_elementarycontractions2} \\
-\bar{\kappa}^{\ast}_{k_1k_2}(l,r) &=   \sum_{s,t=0}^\infty\frac{1}{s!t!}\langle \Phi|\textbf{L}^{s} \beta^\dagger_{k_2}\beta^\dagger_{k_1} \textbf{R}^{t}|\Phi\rangle_c  \nonumber \\
& \equiv  \sum_{s,t=0}^\infty \langle _s | ^{k_2} | \, |^{k_1} |^t  \rangle_c  \,, \label{expansion_elementarycontractions3} \\
-\sigma^{\ast}_{k_1k_2}(l,r) &=  \sum_{s,t=0}^\infty\frac{1}{s!t!}\langle \Phi|\textbf{L}^{s} \beta_{k_2}\beta^\dagger_{k_1} \textbf{R}^{t}|\Phi\rangle_c  \nonumber \\
& \equiv  \sum_{s,t=0}^\infty \langle _s | _{k_2} | \, |^{k_1} |^t  \rangle_c  \, , \label{expansion_elementarycontractions4}
\end{align}
\end{subequations}
knowing that the diagrammatic rules applicable to the four connected\footnote{The connected character of the presently introduced matrix elements is defined with respect to the two operators $\beta^{(\dagger)}_{k_2}$ and $\beta^{(\dagger)}_{k_1}$ that translate diagrammatically into two external lines; see \ref{diagrules} for details.} matrix elements introduced through Eq.~\eqref{expansion_elementarycontractions} are laid out in~\ref{diagrules}.
\end{enumerate}

Taking into account all multiplicative factors pointed out above and summing over all allowed numbers $k$ of normal strings/contractions\footnote{The sum over $k$ starts from 0 (1) and runs over even (odd) integers whenever $i$ and $j$ are even (odd).}, one eventually obtains Eq.~\eqref{eq:gen_Oij_before} under the final form
\begin{align}\label{eq:gen_kern_tot}
  \frac{\langle  \Phi(l) | \textbf{O}^{ij} |  \Phi(r) \rangle}{\langle  \Phi(l) |  \Phi(r) \rangle} &=\sum_{\substack{k=0\\\text{parity}}}^{\text{min}(i,j)} \!\!c(i,j,k) \, \textbf{o}^{k_1 \ldots k_i}_{l_1 \ldots l_j} \nonumber \\
  &\hspace{0.8cm} \times \rho_{l_1k_1}(l,r) \ldots \rho_{l_{k}k_{k}}(l,r) \nonumber \\
  &\hspace{0.8cm} \times \bar{\kappa}^{\ast}_{k_{k+1}k_{k+2}}(l,r) \ldots \bar{\kappa}^{\ast}_{k_{i-1}k_{i}}(l,r) \nonumber \\
  &\hspace{0.8cm} \times \kappa_{l_{k+1}l_{k+2}}(l,r) \ldots  \kappa_{l_{j-1}l_{j}}(l,r)\,,
\end{align}
which proves the off-diagonal Wick's theorem~\cite{Balin69wick} by expressing the connected operator kernel in terms of off-diagonal elementary contractions. Applying Eq.~\eqref{eq:gen_kern_tot} to the sum of $\textbf{O}^{ij}$ operators characterized by $n\neq 3$, the expression of the connected operator kernel of a three-body operator ($N=3$) is given in \ref{examplerank3operator}.

While the ODWT has indeed been formally recovered, a full proof requires an explicit computation of the off-diagonal elementary contractions themselves via the symmetric approach. As clearly illustrated in Eq.~\eqref{expansion_elementarycontractions}, each of these contractions takes itself the form of an infinite, non-terminating, expansion. As demonstrated below, these expansions however happen to deliver known power series that are shown to be equal to the expressions obtained via the asymmetric approach in \ref{standardGWT}. 

\subsubsection{Elementary contractions}
\label{sec:O11}

\begin{figure*}
    \includegraphics[width=1.0\textwidth]{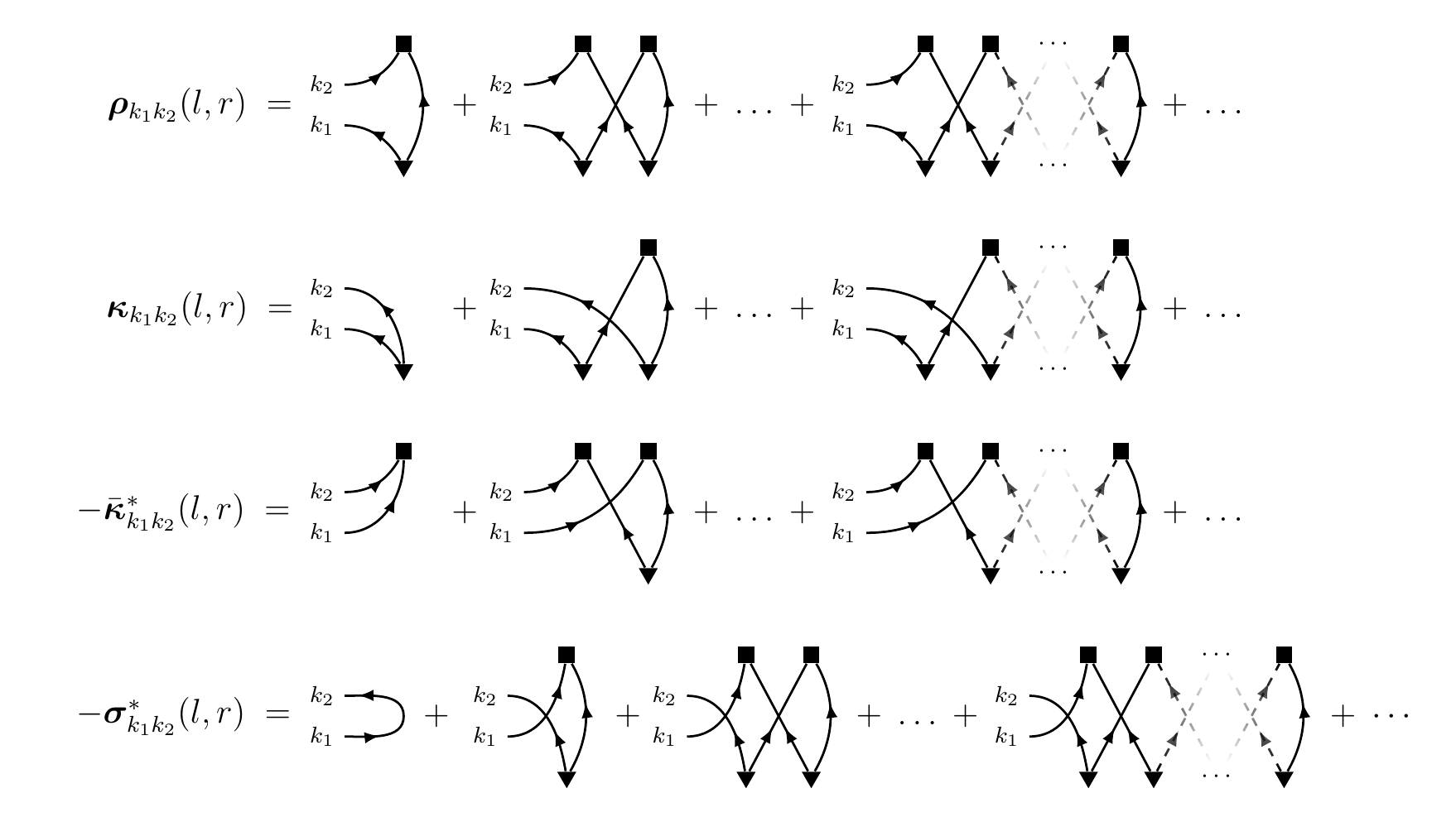}
\caption{
\label{rhodiag}
Diagrammatic expansions of the four off-diagonal elementary contractions.}
\end{figure*}

The first elementary contraction $\rho_{k_1k_2}(l,r)$ (Eq.~\eqref{expansion_elementarycontractions1}) sums the connected matrix elements $\langle _s | ^{k_2} | \, |_{k_1} |^t  \rangle_c$ over all $s$ and $t$ values, knowing in fact that both integers are constrained to be equal, i.e. $s=t\equiv n$. The matrix element $\langle _n | ^{k_2} | \, |_{k_1} |^n  \rangle_c$ is made out of a single unlabelled connected diagram with two external legs. Indeed, there is only one topologically distinct way to connect $\beta^{\dagger}_{k_2}$ to $\beta_{k_1}$ via an alternate succession of $n$ operators $\textbf{L}$ and $n$ operators $\textbf{R}$. By virtue of Eq.~\eqref{generalizeddensitymatrix3}, the term of order $n=0$ is zero in the present case. 

The corresponding diagrammatic expansion of $\rho_{k_1k_2}(l,r)$ is provided in Fig.~\ref{rhodiag}. Diagrammatic rules deliver the algebraic expressions for each order $n$, thus leading to
\begin{strip}
\begin{align}
\label{eq:final1stcontraction}
\rho_{k_1k_2}(l,r) &= + \sum_{n=1}^\infty
    \sum_{\substack{h_1h_2\\...\\h_{2n-1}}}\textbf{z}^{k_1h_1}(r)\textbf{z}^*_{h_1h_2}(l)\textbf{z}^{h_2h_3}(r)\ldots\textbf{z}^*_{h_{2n-3}h_{2n-2}}(l)\textbf{z}^{h_{2n-2}h_{2n-1}}(r)\textbf{z}^*_{k_2h_{2n-1}}(l)\nonumber\\
    &= -\sum_{n=0}^\infty
    \Big[\textbf{z}(r)\Big(\textbf{z}^*(l)\textbf{z}(r)\Big)^n\textbf{z}^*(l)\Big]_{k_1k_2} \nonumber\\
    &= -\left[\textbf{z}(r)\frac{1}{1-\textbf{z}^{*}(l)\textbf{z}(r)}\textbf{z}^*(l)\right]_{k_1k_2} \,,
\end{align}
\end{strip}%
where the Taylor series
\begin{equation}
    \sum_{n=0}^{\infty}\varepsilon^n = \frac{1}{1-\varepsilon} \label{PS1}
\end{equation}
has been used to resum the infinite expansion.

The three other elementary contractions (Eqs.~(\ref{expansion_elementarycontractions2}- \ref{expansion_elementarycontractions4})) can be calculated similarly. Their diagrammatic expansions are also displayed in Fig.~\ref{rhodiag}, where one notices that the fourth contraction contains a non-zero term of order $n=0$. The corresponding algebraic expressions are given by
\begin{subequations}
\label{offdiagelemcontract}
\begin{align}
\kappa_{k_1k_2}(l,r) &= +\sum_{n=0}^\infty
    \Big[\textbf{z}(r)\Big(\textbf{z}^*(l)\textbf{z}(r)\Big)^n\Big]_{k_1k_2} \nonumber\\
    &= +\left[\textbf{z}(r)\frac{1}{1-\textbf{z}^{*}(l)\textbf{z}(r)}\right]_{k_1k_2} \,, \label{eq:3rdelementarycontraction} \\
-\bar{\kappa}^{\ast}_{k_1k_2}(l,r) &= -\sum_{n=0}^\infty
    \Big[\Big(\textbf{z}^*(l)\textbf{z}(r)\Big)^n\textbf{z}^*(l)\Big]_{k_1k_2} \nonumber\\
    &= -\left[\frac{1}{1-\textbf{z}^{*}(l)\textbf{z}(r)}\textbf{z}^*(l)\right]_{k_1k_2} \, , \label{eq:2ndelementarycontraction}\\
-\sigma^{\ast}_{k_1k_2}(l,r) &= +\sum_{n=0}^\infty
    \Big[\Big(\textbf{z}^*(l)\textbf{z}(r)\Big)^n\Big]_{k_1k_2} \nonumber\\
    &= +\left[\frac{1}{1-\textbf{z}^{*}(l)\textbf{z}(r)}\right]_{k_1k_2} \, . \label{eq:4ndelementarycontraction}
\end{align}
\end{subequations}
It is easy to check that the four properties listed in Eqs.~\eqref{densitymatrices} are indeed satisfied by the off-diagonal contractions. 

Eventually, the off-diagonal contractions have been  expressed as a known power series in the variable $\textbf{z}^{*}(l)\textbf{z}(r)$. While this result looks very different from the one obtained from the asymmetric approach where the off-diagonal contractions read as linear functions of the Thouless matrix $\textbf{z}(l,r)$, it can be easily shown, as stipulated in \ref{standardGWT}, that both sets of expressions are in fact identical.

\subsection{Norm kernel}

The norm kernel reads in the symmetric approach as
\begin{align}
\label{eq:normkernel_diag}
\frac{\langle \Phi(l) |  \Phi(r) \rangle}{\langle \Phi(l) |\Phi \rangle \langle \Phi | \Phi(r) \rangle} &=   \sum_{s,t=0}^\infty\frac{1}{s!t!}\langle \Phi| \textbf{L}^{s} \textbf{R}^{t}|\Phi\rangle  \nonumber  \\
&\equiv \sum_{s,t=0}^\infty \langle _s | \,  |^t  \rangle  \,,
\end{align}
where the condition $s=t\equiv n$ must be fulfilled.

\subsubsection{Exponentiation of closed diagrams}

\begin{figure*}
   \includegraphics[width=1.0\textwidth]{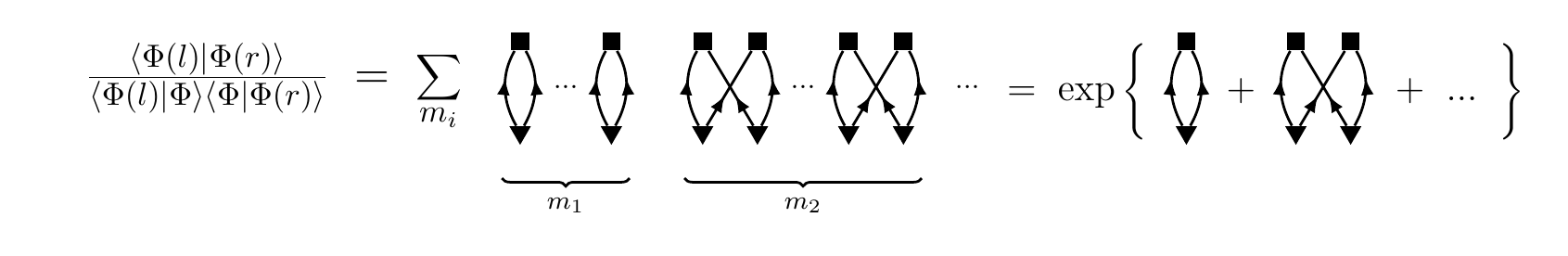}
\caption{
\label{diagramnormkernel}
Representation of the norm kernel in terms of (repeated) disjoint closed diagrams and as the exponential of the sum of distinct closed diagrams.}
\end{figure*}

As already discussed, and as detailed in \ref{diagrules}, diagrams contributing to $\langle _n | \,  |^n  \rangle$ are composed of disjoint closed sub-diagrams. A  closed\footnote{The trivial diagram obtained for $n=0$ is $\langle _0 | \,  |^0  \rangle = \langle \Phi | \Phi \rangle=1$. Since it contains neither vertices nor lines, it does not qualify as a closed diagram.} diagram involves an equal number of successively connected operators $\textbf{L}$ and $\textbf{R}$. For a given $n_i \geq 1$, there exists in fact a single topologically distinct unlabelled closed diagram\footnote{In the present discussion $\Gamma_{cl}(n_i)$ equally represents the closed diagram and its algebraic contribution.}
\begin{align}
    \Gamma_{cl}(n_i)&\equiv \langle _{n_i} | \,  |^{n_i}  \rangle_{cl}\,, \label{closed_contrib0}
\end{align}
where the index $cl$ stands for {\it closed}. 

A generic diagram $\Gamma(n)$ contributing to $\langle _n | \,  |^n  \rangle$ factorizes into $m_1$ identical closed subdiagrams $\Gamma_{cl}(n_1)$, $m_2$  identical closed sub-diagrams $\Gamma_{cl}(n_2)$, and so on. Obviously, a closed diagram $\Gamma_{cl}(n_i)$ can only contribute whenever $n\geq n_i$. Using the convention that a closed diagram occurs with multiplicity $m_i=0$ whenever $n<n_i$, a diagram contributing to $\langle _n | \,  |^n  \rangle$ can be written as
\begin{equation}
    \Gamma(n)=\frac{[\Gamma_{cl}(n_1)]^{m_1}}{m_1!}\frac{[\Gamma_{cl}(n_2)]^{m_2}}{m_2!}\cdots \label{contributionorder_n_norm}
\end{equation}
where the product runs over all possible closed strings $\Gamma_{cl}(n_i)$ such that the condition
\begin{equation}
    \sum_{n_i=1}^{\infty} m_i n_i=n\,, \label{constraintorder}
\end{equation}
is satisfied. According to the diagrammatic rules, the symmetry factor of $\Gamma(n)$ must be worked out. In addition to the symmetry factors associated with each closed subdiagram (included into $\Gamma_{cl}(n_i)$), the symmetry factor $S_{\Gamma(n)}$ incorporates the set of denominators in Eq.~\eqref{contributionorder_n_norm}. Each such denominator $m_i!$ denotes the number of permutations of the equivalent groups of $\textbf{L}$ and $\textbf{R}$ operators belonging to the $m_i$ identical closed sub-diagrams $\Gamma_{cl}(n_i)$.

Summing over all topologically distinct diagrams $\Gamma(n)$, with $n$ running from $0$ to $\infty$, all possible closed contributions $\Gamma_{cl}(n_i)$ eventually occur with arbitrary integer multiplicities $m_i$ such that
\begin{align}
    \frac{\langle \Phi(l) | \Phi(r) \rangle}{\langle \Phi(l) | \Phi \rangle \langle \Phi | \Phi(r) \rangle}  &=  \sum_{n=0}^{\infty} \sum_{\Gamma(n)} \Gamma(n)\nonumber\\
    &= \sum_{m_1m_2\ldots =0}^{\infty} \frac{[\Gamma_{cl}(1)]^{m_1}}{m_1!}\frac{[\Gamma_{cl}(2)]^{m_2}}{m_2!}\cdots\nonumber\\
    &= \exp\Big\{\sum_{n=1}^{\infty}\Gamma_{cl}(n)\Big\}\,, \label{expofclosed}
\end{align}
thus demonstrating the exponentiation of closed diagrams in the expansion of the norm kernel. The above rationale is illustrated diagrammatically in Fig.~\ref{diagramnormkernel}.

\begin{figure*}
    \centering
   \includegraphics[width=0.9\textwidth]{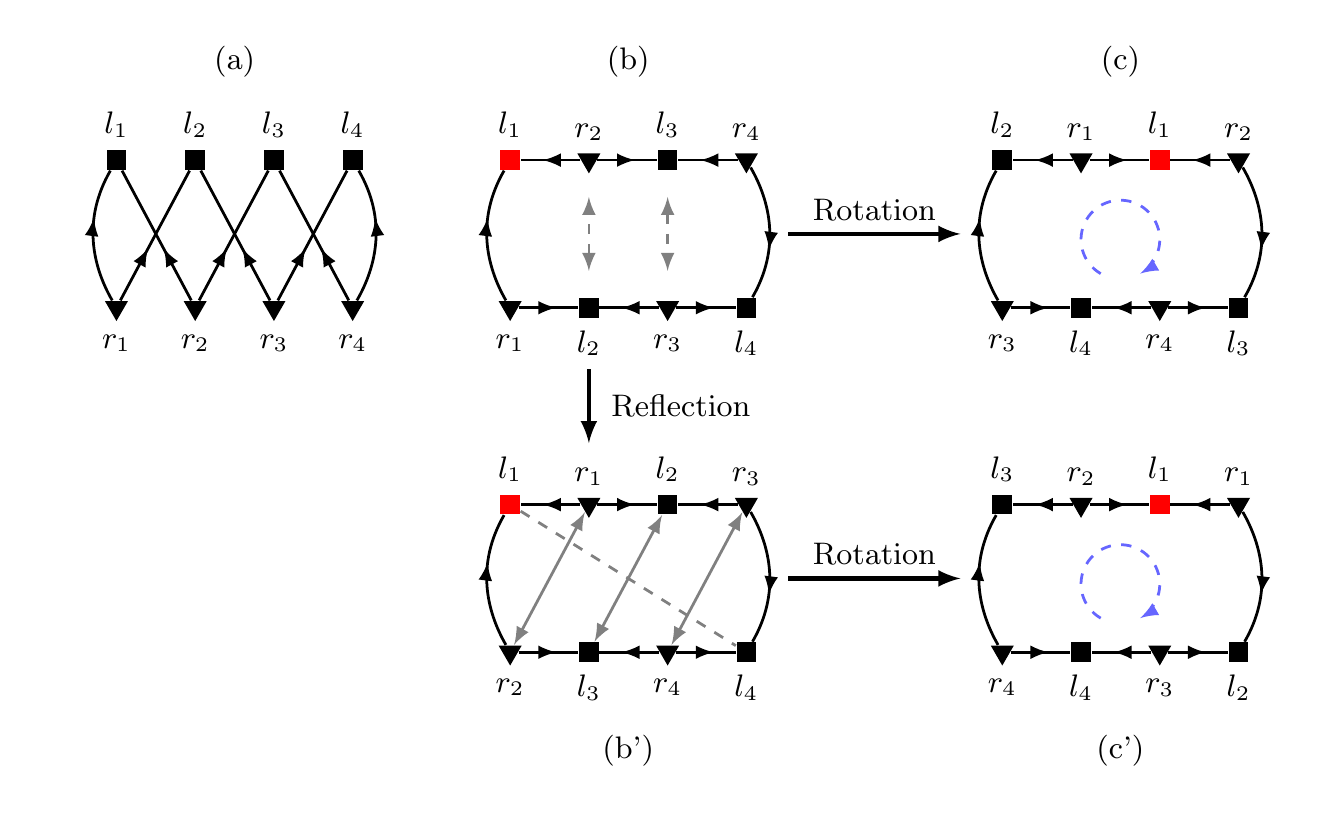}
\caption{
\label{symmetryfactor}
Permutations of the labelled diagram $\Gamma_{cl}(4)$ delivering topologically equivalent diagrams. (a): Labelled diagram in the original representation. (b): Same diagram but expanded in a way that makes permutations of the vertices more transparent. (b'): Labelled diagram obtained from (b) via a reflection with respect to the diagonal $l_1-l_4$ inverting the clockwise ordering of the labelled vertices. (c) and (c'): one representative of the four clockwise circular permutations obtained from (b) and (b'), respectively. There is thus a total of $S_{\Gamma_{cl}(4)}=2\times 4$ permutations out of which the original labelled diagram can be recovered by a mere translation of the vertices in the plane.}
\end{figure*}

\subsubsection{Resummation of closed diagrams}

One is now left with the computation of the exponent in Eq.~\eqref{expofclosed}. According to the diagrammatic rules, the algebraic expression of the single topologically distinct unlabelled closed diagram of order $n$ reads as
\begin{align}\label{eq:norm_term}
\Gamma_{cl}(n)&=  -\frac{1}{2n}  \text{Tr}\Big\{\Big(\textbf{z}^*(l)\textbf{z}(r)\Big)^{n}\Big\}  \, ,
\end{align}
where the symmetry factor is $S_{\Gamma_{cl}(n)}=2n$. As visible from Fig.~\ref{diagramnormkernel}, the $1/2$ factor obtained for $n=1$ relates to the existence of a pair of equivalent lines. For $n>1$, the symmetry factor relates to the number of permutations of the $\textbf{L}$ and $\textbf{R}$ vertices in the {\it labelled} diagram $\Gamma_{cl}(n)$ delivering topologically equivalent labelled diagrams. As illustrated in Fig.~\ref{symmetryfactor} for $n=4$, there are $S_{\Gamma_{cl}(n)}=2n$ such permutations, made out of the convolution of
\begin{enumerate}
    \item The identity plus $1$ reflection inverting the clockwise ordering,
    \item $n$ clockwise circular permutations.
\end{enumerate}

Summing over all closed diagrams, the final result is obtained as
\begin{align}
\label{eq:norm_term2}
\sum_{n=1}^{\infty} \Gamma_{cl}(n)    
&= - \frac{1}{2} \text{Tr}\Big\{ \sum_{n=1}^\infty \frac{1}{n} \Big(\textbf{z}^*(l)\textbf{z}(r)\Big)^{n}\Big\}  \nonumber\\
&= \frac{1}{2}\text{Tr}\Big\{\ln\big(1-\textbf{z}^{*}(l)\textbf{z}(r)\big)\Big\} \,,
\end{align}
where the Taylor series 
\begin{equation}
- \sum_{n=1}^\infty \frac{\varepsilon^n}{n} = \ln(1-\varepsilon)\,,    \label{taylorserieslog}
\end{equation}
has been used to resum the infinite expansion.


\subsubsection{Onishi formula}

Combining Eqs.~\eqref{expofclosed} and~\eqref{eq:norm_term2}, the norm kernel eventually reads as
\begin{equation}\label{eq:Onishi}
\frac{\langle \Phi(l) | \Phi(r) \rangle}{\langle \Phi(l) |\Phi \rangle \langle \Phi | \Phi(r) \rangle} =  \exp\Big[\frac{1}{2}\text{Tr}\Big\{\ln\big(1-\textbf{z}^{*}(l)\textbf{z}(r)\big)\Big\}\Big]\,,
\end{equation}
which is nothing but the well-celebrated Onishi formula~\cite{onishi66}. It has often been stated that the Onishi formula\footnote{Strickly speaking, the Onishi formula, as any formula based on the symmetric approach, can only deliver the phase of the overlap $\langle \Phi(l) | \Phi(r) \rangle$ modulo the knowledge of the phase associated with, i.e. initially fixed for, the overlaps $\langle \Phi(l) |\Phi \rangle$ and $\langle \Phi|\Phi(r) \rangle$ of the two involved states with respect to the reference Bogoliubov state $|\Phi \rangle$.} is compromised by an undefined complex phase. In the derivation above based on the application of standard Wick's theorem, there is no algebraic manipulation that can be responsible for a loss of phase. Thus, the loss of phase due to the apparent necessity to compute the square root originating from the factor $1/2$ in the exponent in Eq.~\eqref{eq:Onishi} can only be fictitious. The same observation is at the heart of Ref.~\cite{Mizusaki:2017ist}. The key point relates to the fact that the eigenvalues of the matrix $\textbf{z}^{*}(l)\textbf{z}(r)$, which is the product of two skew-symmetric matrices, are doubly-degenerate~\cite{neergard83a,Oi:2019yrw}. This double degeneracy necessarily compensates for the incriminated factor $1/2$ in Eq.~\eqref{eq:Onishi}, thus demonstrating the fictitious character of the square root and of the apparent loss of phase~\cite{Bally:2017nom,Mizusaki:2017ist}. 

\subsubsection{Pfaffian formula}

The equivalence of the Onishi and Pfaffian~\cite{Robledo:2009yd} formulae for the norm kernel has been demonstrated in a pedestrian way in Ref.~\cite{Mizusaki:2017ist}. As a matter of fact, the two formulae can be directly connected by exploiting the generic identity~\cite{krivoruchenko2016trace}
\begin{equation}
    \text{pf}(A)\text{pf}(B)=\exp\left[\frac{1}{2}\text{Tr}\ln(A^{\text{T}}B)\right]\, , \label{identitypfaffian}
\end{equation}
where $A$ and $B$ are two skew-symmetric matrices such that $A^TB$ is itself a positive-definite matrix. Rewriting the argument of the logarithm in Eq.~\eqref{eq:Onishi} as
\begin{align}
    1-\textbf{z}^{*}(l)\textbf{z}(r)=\textbf{z}^{*}(l)\big(\textbf{z}^{*}(l)^{-1}-\textbf{z}(r)\big)\, , 
\end{align}
Eq.~\eqref{identitypfaffian} is applied for 
\begin{subequations}
\begin{align}
    A&\equiv \textbf{z}^\dagger(l)\\
    B&\equiv \textbf{z}^{*}(l)^{-1}-\textbf{z}(r)\,.
\end{align}
\end{subequations}
This leads to expressing the norm overlap as%
\begin{align}\label{eq:pf_prod}
\frac{\langle \Phi(l) | \Phi(r) \rangle}{\langle \Phi(l) |\Phi \rangle \langle \Phi | \Phi(r) \rangle} &= \text{pf}\big(\textbf{z}^\dagger(l)\big)\text{pf}\big(\textbf{z}^{*}(l)^{-1}-\textbf{z}(r)\big)\nonumber\\
    &= (-1)^{n_1/2}\text{pf}\big(\textbf{z}^*(l)\big)\text{pf}\big(\textbf{z}^{*}(l)^{-1}+\textbf{z}^{\text{T}}(r)\big)\,,\nonumber%
\end{align}%
where the last equivalence makes use of the property %
\begin{equation}
    \text{pf}(A^{\text{T}})=(-1)^{n}\text{pf}(A)\,,
\end{equation}%
with $2n$ the size of the matrix $A$, i.e. the (even) dimension $n_1$ of the (truncated) one-body Hilbert space $\mathcal{H}_1$ in the present context. The last useful identity for the pfaffian of a skew-symmetric matrix displaying the structure
\begin{equation}
    S=\begin{pmatrix}
    M&Q\\
    -Q^T&N
    \end{pmatrix}\,,
\end{equation}
with $M$ an invertible matrix, is given by~\cite{caianiello1973combinatorics}
\begin{equation}
    \text{pf}(S)=\text{pf}(M)\text{pf}(N+Q^{\text{T}}M^{-1}Q)\,.
\end{equation}
Identifying
\begin{subequations}
\begin{align}
    M&\equiv \textbf{z}^*(l)\,,\\
    N&\equiv \textbf{z}^{\text{T}}(r)\,,\\
    Q&\equiv 1\,,
\end{align}
\end{subequations}
the pfaffian formulation of the norm overlap
\begin{equation}
\frac{\langle \Phi(l) | \Phi(r) \rangle}{\langle \Phi(l) |\Phi \rangle \langle \Phi | \Phi(r) \rangle} = (-1)^{n_1/2}\text{pf}\,\bigg[\begin{pmatrix}
    \textbf{z}^*(l) & 1\\
    -1 & \textbf{z}^{\text{T}}(r)
    \end{pmatrix}\bigg]\,,
\end{equation}
is eventually obtained from the Onishi formula, thus bypassing the intermediate apparent phase undetermination.

\section{Discussion and conclusions}
\label{conclusions}

The interest of the present work is primarily formal and conceptual. The main goal has been to offer a novel perspective on the off-diagonal Wick theorem and the Onishi formula by consistently computing the off-diagonal operator and norm kernels at play in, e.g., the projected generator coordinate method via a single formal approach. The method expresses the two Bogoliubov states at play with respect to a third reference state $| \Phi \rangle$ via Thouless' theorem such that the kernels of interest read as
\begin{subequations}
\begin{align}
\frac{\langle \Phi(l) | O | \Phi(r) \rangle}{\langle \Phi(l) | \Phi(r) \rangle} & =  \frac{\langle \Phi | e^{\textbf{Z}^{20}(l)^\dagger} O e^{\textbf{Z}^{20}(r)} |\Phi \rangle}{\langle \Phi | e^{\textbf{Z}^{20}(l)^\dagger} e^{\textbf{Z}^{20}(r)} |\Phi \rangle}  \, , \label{kernel1con} \\
\frac{\langle \Phi(l) | \Phi(r) \rangle}{\langle \Phi(l)|\Phi \rangle \langle \Phi|\Phi(r) \rangle} &\equiv  \langle \Phi | e^{\textbf{Z}^{20}(l)^\dagger} e^{\textbf{Z}^{20}(r)} |\Phi \rangle \, .
\end{align}
\end{subequations}
While the exponentials of the Thouless one-body operators $\textbf{Z}^{20}(l)^\dagger$ and $\textbf{Z}^{20}(r)$ do not naturally terminate, a diagrammatic method was used to demonstrate that the infinite set of terms can be resummed exactly. 

Interestingly, the diagrammatic technique and the associated infinite resummations leading to the exact computation of the above kernels can be used to design non-trivial approximations to more complex kernels of interest. As the simplest example, replacing the two Thouless one-body operators by a two-body cluster amplitude into Eq.~\eqref{kernel1con} leads to
\begin{align}
E^{\text{vCCD}} & =  \frac{\langle \Phi | e^{T_2^\dagger} H e^{T_2} |\Phi \rangle}{\langle \Phi | e^{T_2^\dagger} e^{T_2} |\Phi \rangle}  \, , \label{kernelvCCD}
\end{align}
which is nothing but the energy at play in the variational coupled cluster with doubles method~\cite{bartlett88a}. Because $T_2$ is now a two-body operator, the equivalent to the off-diagonal Wick theorem and the associated resummation of the infinite expansion of the exponentials do not hold. As a result, approximation schemes have to be set up~\cite{robinson12a,marie21a} such that designing more advanced truncation schemes than existing ones can be of interest. Because the diagrammatic does not require the operators in the two exponentials to be the same, addressing even more general kernels (including the corresponding norm kernels) than the one displayed in Eq.~\eqref{kernelvCCD} is envisioned.

\section*{Acknowledgements}

A.P. is supported by the CEA NUMERICS program, which has received funding from the European Union's Horizon 2020 research and innovation program under the Marie Sk{\l}odowska-Curie grant agreement No 800945. 

\section*{Data Availability Statement}
This manuscript has no associated data or the data will not be deposited.

\begin{appendix}

\section{Normal-ordered operator}
\label{NO_operators}

An arbitrary rank-$N$ particle-number-conserving fermionic operator \(O\) can be written as
\begin{equation} 
	O\equiv\sum_{n=0}^N {O^{nn}} \, , \label{generalop}
\end{equation}
where
\begin{equation} 
	O^{mn}\equiv \frac{1}{m!} \frac{1}{n!}  
	\sum_{\substack{a_1\cdots a_m\\b_1\cdots b_n}}
	o^{a_1\cdots a_m}_{b_1\cdots b_n} \, 
	c^\dag_{a_1}\cdots c^\dag_{a_m}c_{b_n}\cdots c_{b_1} \, , \label{defnbodyopvacuum}
\end{equation}
contains $m(n)$ particle creation (annihilation) operators. The zero-body part $O^{00}$ is the scalar obtained as the expectation value of $O$ in the particle vacuum
\begin{equation}
    O^{00}=\langle 0|O|0 \rangle \, .
\end{equation}
In Eq.~\eqref{defnbodyopvacuum},  matrix elements are fully anti-symmetric under the exchange of any pair of upper or lower indices, i.e.
\begin{equation}
  o^{a_1\cdots a_m}_{b_1\cdots b_n} = \epsilon(\sigma_u) \epsilon(\sigma_l)  \, o^{\sigma_u(a_1\cdots a_m)}_{\sigma_l(b_1\cdots b_n)} \, ,
\end{equation}
where $\epsilon(\sigma_u)$  ($\epsilon(\sigma_l)$) refers to the signature of the permutation $\sigma_u(\ldots)$ ($\sigma_l(\ldots)$) of the $m$ ($n$) upper (lower) indices. In case the particle-number conserving operator is hermitian, each term $O^{nn}$ is hermitian with its matrix elements fulfilling
\begin{equation}
o^{a_1\cdots a_n}_{b_1\cdots b_n} = \left(o_{a_1\cdots a_m}^{b_1\cdots b_n}\right)^\ast  \, .
\end{equation}

By virtue of standard Wick's theorem~\cite{Wick50theorem}, the operator $O$ can be normal ordered with respect to the Bogoliubov vacuum $| \Phi \rangle$
\begin{align}
O&\equiv\sum_{n=0}^N\textbf{O}^{[2n]} \equiv\sum_{n=0}^N\sum_{\substack{i,j=0\\i+j=2n}}^{2n}\textbf{O}^{ij}
\end{align}
where the component
\begin{equation}
    \textbf{O}^{ij}\equiv\frac{1}{i!j!}\sum_{\substack{k_1...k_i\\l_1...l_j}}\textbf{o}^{k_1...k_i}_{l_1...l_j}\beta^\dagger_{k_1}...\beta^\dagger_{k_i}\beta_{l_j}...\beta_{l_1} \label{componentNO_op}
\end{equation}
contains $i(j)$ quasi-particle creation (annihilation) operators. The zero-body part $\textbf{O}^{[0]}$ is the scalar obtained as the expectation value of $O$ in the Bogoliubov vacuum%
\begin{equation}
    \textbf{O}^{00}= \langle \Phi|O|\Phi \rangle \, .
\end{equation}
In Eq.~\eqref{componentNO_op}, matrix elements are fully anti-symmetric under the exchange of any pair of upper or lower indices, i.e.
\begin{equation}
  {\mathbf o}^{k_1\cdots k_i}_{l_1\cdots l_j} = \epsilon(\sigma_u) \epsilon(\sigma_l)  \, {\mathbf o}^{\sigma_u(k_1\cdots k_i)}_{\sigma_l(l_1\cdots l_j)} \, . \label{antisymME}
\end{equation}
These matrix elements are functionals of the Bogoliubov matrices $(\mathcal{U},\mathcal{V})$ associated with  $| \Phi \rangle$ and of the matrix elements $\{o^{a_1\cdots a_n}_{b_1\cdots b_n}\}$ initially defining the operator $O$. As such, the content of each operator ${\mathbf O}^{ij}$ depends on the rank $N$ of $O$. For more details about the normal ordering procedure and for explicit expressions of the matrix elements up to $N=3$, see Refs.~\cite{Duguet:2015yle,Tichai18BMBPT,Arthuis:2018yoo,Ripoche2020}.

In case the operator is hermitian, each component $\textbf{O}^{[2n]}$ is itself hermitian with $\textbf{O}^{ij} = \textbf{O}^{ji\dagger}$ such that matrix elements satisfy
\begin{equation}
{\mathbf o}^{k_1\cdots k_i}_{l_1\cdots l_j} = \left({\mathbf o}^{l_1\cdots l_j}_{k_1\cdots k_i}\right)^\ast \, .
\end{equation}

\section{Diagrammatic rules}
\label{diagrules}

The present appendix is dedicated to setting up the diagrammatic rules allowing one to compute matrix elements of the form 
\begin{subequations}
\label{basicME}
\begin{align}
\langle _s|^i \, _j| ^t \rangle &\equiv \frac{1}{s!} \frac{1}{t!} \langle \Phi | \textbf{L}^s \textbf{O}^{ij}  \textbf{R}^t  |\Phi \rangle \, , \label{basicME1} \\
\langle _s| \,|^t \rangle &\equiv \frac{1}{s!} \frac{1}{t!} \langle \Phi | \textbf{L}^s   \textbf{R}^t  |\Phi \rangle \, , \label{basicME2} 
\end{align}
\end{subequations}
where $\textbf{O}^{ij}$ takes the form given in Eq.~\eqref{componentNO_op} and where  $\textbf{R}$ ($\textbf{L}$) is a one-body\footnote{The diagrammatic rules can be straightforwardly generalized to operators $\textbf{R}$ and $\textbf{L}$ of higher, and possibly different, ranks.} excitation (de-excitation) operator as defined in Eq.~\eqref{defR} (Eq.~\eqref{defL}). 

The diagrammatic rules are also worked out to compute a second category of matrix elements of present interest%
\begin{subequations}
\label{basiccont}
\begin{align}
\langle _s | ^{k_2} | \, |_{k_1} |^t  \rangle &\equiv \frac{1}{s!} \frac{1}{t!} \langle \Phi | \textbf{L}^s \beta^{\dagger}_{k_2} \beta_{k_1} \textbf{R}^t  |\Phi \rangle \, , \label{basiccont1}\\
\langle _s | _{k_2} | \, |_{k_1} |^t  \rangle &\equiv \frac{1}{s!} \frac{1}{t!} \langle \Phi | \textbf{L}^s \beta_{k_2} \beta_{k_1}  \textbf{R}^t  |\Phi \rangle \, , \label{basiccont2}\\
\langle _s | ^{k_2} | \, |^{k_1} |^t  \rangle &\equiv \frac{1}{s!} \frac{1}{t!} \langle \Phi | \textbf{L}^s \beta^{\dagger}_{k_2} \beta^{\dagger}_{k_1} \textbf{R}^t  |\Phi \rangle \, , \label{basiccont3}\\
\langle _s | _{k_2} | \, |^{k_1} |^t  \rangle &\equiv \frac{1}{s!} \frac{1}{t!} \langle \Phi | \textbf{L}^s \beta_{k_2} \beta^{\dagger}_{k_1}  \textbf{R}^t  |\Phi \rangle \, . \label{basiccont4}
\end{align}
\end{subequations}
These matrix elements differ from the two introduced in Eq.~\eqref{basicME} by the presence of two "external/fixed" quasi-particle operators, i.e. quasi-particle operators whose indices are not summed over.

The diagrammatic rules derive from the straight application of standard Wick's theorem with respect to the Bogoliubov vacuum $|\Phi \rangle$. The application of Wick's theorem delivers the complete set of fully contracted terms associated with  the operator product entering the matrix element of interest. Given that the operators at play are all conveniently expressed in the quasi-particle basis associated with the Bogoliubov vacuum $| \Phi \rangle$, the four involved elementary contractions take the simplest possible form
\begin{align}
\mathbf{R}_{k_1k_2} &=
\begin{pmatrix}
\frac{\langle \Phi |\, \beta^{\dagger}_{k_2} \beta^{\phantom{\dagger}}_{k_1} | \Phi \rangle}{\langle \Phi | \Phi \rangle} & \frac{\langle \Phi |\, \beta^{\phantom{\dagger}}_{k_2} \beta^{\phantom{\dagger}}_{k_1} | \Phi \rangle}{\langle \Phi | \Phi \rangle} \\
\frac{\langle \Phi |\, \beta^{\dagger}_{k_2} \beta^{\dagger}_{k_1} | \Phi \rangle}{\langle \Phi | \Phi \rangle} &  \frac{\langle \Phi |\, \beta^{\phantom{\dagger}}_{k_2} \beta^{\dagger}_{k_1} | \Phi \rangle}{\langle \Phi | \Phi \rangle}
\end{pmatrix} \nonumber \\
&\equiv
\begin{pmatrix}
R^{+-}_{k_1k_2} & R^{--}_{k_1k_2} \\
R^{++}_{k_1k_2} &  R^{-+}_{k_1k_2}
\end{pmatrix} \nonumber \\
&=
\begin{pmatrix}
0 & 0 \\
0 &  \delta_{k_1k_2}
\end{pmatrix} \ , \label{generalizeddensitymatrix2}
\end{align}
such that the sole non-zero contraction $R^{-+}_{k_1k_2} = \delta_{k_1k_2}$ needs to be considered. 

\subsection{Diagrammatic representation}
\label{subs:diag_represent}

The diagrammatic representation of the various contributions to the matrix elements of interest relies on the definition of the following building blocks
\begin{enumerate}
\item 
As illustrated in Fig.~\ref{vertexOij}, the normal-ordered operator $\textbf{O}^{ij}$ entering the matrix element $\langle _s|^i \, _j| ^t \rangle$ is represented by a Hugenholtz vertex with $i$ ($j$) lines traveling out of (into) it and representing quasi-particle creation (annihilation) operators. 
\begin{figure}
\centering
   \includegraphics[width=.35\textwidth]{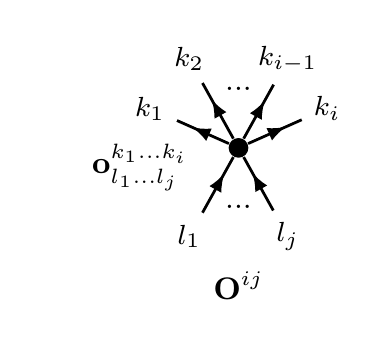}
\caption{
\label{vertexOij}
Diagrammatic representation of the normal-ordered operator $\textbf{O}^{ij}$ as a fully anti-symmetric Hugenholtz vertex.}
\end{figure}
The factor $\textbf{o}^{k_1 \ldots k_i}_{l_1 \ldots l_j}$ is associated to the vertex while assigning indices $k_1 \ldots k_i$ consecutively from the leftmost to the rightmost line above the vertex and indices $l_{1} \ldots l_{j}$ consecutively from the leftmost to the rightmost line below the vertex.
\item As illustrated in Fig.~\ref{vertexLR}, the operator $\textbf{L}$ ($\textbf{R}$) entering all matrix elements of present interest is represented by a vertex of the $\textbf{O}^{02}$ ($\textbf{O}^{20}$) with the associated factor $\textbf{z}^{\ast}_{l_1 l_2}(l)$ ($\textbf{z}^{k_1 k_2}(r)$).
\begin{figure}
\centering
   \includegraphics[width=.4\textwidth]{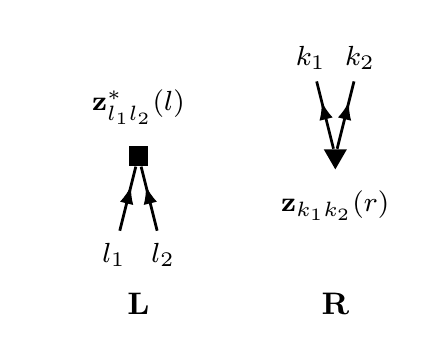}
\caption{
\label{vertexLR}
Diagrammatic representation of the two Thouless operators $\textbf{L}$ (left) and $\textbf{R}$ (right).}
\end{figure}
\item The only non-zero contraction $R^{-+}_{k_1k_2} = \delta_{k_1k_2}$ is represented in Fig.~\ref{f:prop} and connects two up-going lines associated with one annihilation and one creation operator, both carrying the same quasi-particle index. For simplicity, one can eventually represent the contraction as a line carrying a single up-going arrow along with one quasi-particle index.
\begin{figure}
\centering
  \includegraphics[width=.4\textwidth]{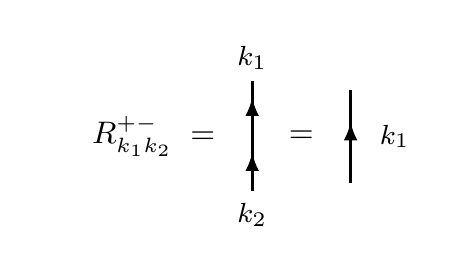}
\caption{
\label{f:prop}
Diagrammatic representation of the single non-zero elementary contraction. The convention is that the left-to-right reading of a matrix element corresponds to the up-down reading of the diagram.}
\end{figure}
\end{enumerate}

\subsection{Diagrams generation}
\label{subs:diag_gene}

With these building blocks at hand, one constructs the diagrams gathering all contributions to the matrix elements introduced in Eqs~\eqref{basicME} and \eqref{basiccont}. The basic rules to do so are as follows
\begin{enumerate}
\item Diagrams contain $s$ square vertices ($\textbf{L}$) and $t$ triangle vertices ($\textbf{R}$), the former being located above the latter.   This is consistent with the convention that the left-to-right reading of a matrix element corresponds to the up-down reading of the diagram.
\item Diagrams making up the two matrix elements introduced in Eqs.~\eqref{basicME} are {\it vacuum-to-vacuum} diagrams with no line leaving the diagram. In $\langle _s|^i \, _j| ^t \rangle$, a dot vertex ($\textbf{O}^{ij}$) is located in between the square and triangle vertices. This is consistent with the convention that the left-to-right reading of a matrix element corresponds to the up-down reading of the diagram.
\item Diagrams making up the four matrix elements introduced in Eqs.~\eqref{basiccont} are {\it linked} with two external lines associated with the operators $\beta^{(\dagger)}_{k_2}$ and $\beta^{(\dagger)}_{k_1}$. The two lines leave the diagram on the same side to the, e.g., left such that (i) both lines are asymptotically in between the square ($\textbf{L}$) and triangle ($\textbf{R}$) vertices and such that (ii) the line carrying index $k_2$ is asymptotically located above the line carrying index $k_1$. This is consistent with the convention that the left-to-right reading of a matrix element corresponds to the up-down reading of the diagram. The arrow carried by each of the two lines points towards the interior (exterior) of the diagram if it is associated with a quasi-particle creation (annihilation) operator.
\item The fact that $R^{-+}_{k_1k_2}$ is the sole non-zero contraction implies that the number of quasi-particle creation operators involved in a given matrix element is equal to the number of quasi-particle annihilation operators. Given that each operator $\textbf{L}$ ($\textbf{R}$) contains two quasi-particle annihilation (creation) operators, this property require the following conditions to be fulfilled
\begin{enumerate}
\item $\langle _s|^i \, _j| ^t \rangle $ demands $t = s + (i-j)/2$\, ,
\item $\langle _s| \, | ^t \rangle $ demands  $t = s$ \, ,
\item $\langle _s | ^{k_2} | \, |_{k_1} |^t  \rangle$ demands $t=s$ \, ,
\item $\langle _s | _{k_2} | \, |_{k_1} |^t  \rangle $ demands $t=s+1$ \, ,
\item $\langle _s | ^{k_2} | \, |^{k_1} |^t  \rangle$ demands $t=s-1$ \, ,
\item $\langle _s | _{k_2} | \, |^{k_1} |^t  \rangle$ demands $t=s$ \, .
\end{enumerate}
such that $t\geq (i-j)/2$, $t\geq 1$ and  $s\geq 1$ in case (a), (d) and (e), respectively.
\item Given the above considerations, one must construct all possible {\it topologically distinct unlabelled} diagrams from the building blocks; i.e., contract together the lines belonging to the $s$ square ($\textbf{L}$) vertices and to the $t$ triangle ($\textbf{R}$) vertices, along with those belonging to the dot ($\textbf{O}^{ij}$) vertex or to the two external ($\beta^{(\dagger)}_{k_2}$ and $\beta^{(\dagger)}_{k_1}$) lines when applicable, in all possible ways. Unlabelled diagrams correspond to diagrams in which $\textbf{L}$ ($\textbf{R}$) vertices are not distinguished by a label.  Topologically distinct unlabelled diagrams cannot be obtained from one another via a mere displacement, i.e. translation, of the vertices in the plane of the drawing.
\item The above process is constrained by the following properties 
\begin{enumerate}
\item Because the operators $\textbf{L}$, $\textbf{R}$, and $\textbf{O}^{ij}$ are in normal-ordered form with respect to $| \Phi \rangle$, self-contractions must be ignored.
\item Because $R^{-+}_{k_1k_2}$ is the sole non-zero contraction, $\textbf{L}$ ($\textbf{R}$) operators cannot display contractions among themselves, i.e. they necessarily contract with $\textbf{R}$ ($\textbf{L}$) operators, along with the $i$ ($j$) quasi-particle creation (annihilation) operators inside $\textbf{O}^{ij}$ or with $\beta^{\dagger}_{k_2}$ ($\beta_{k_2}$) and/or  $\beta^{\dagger}_{k_1}$ ($\beta_{k_1}$) when applicable. 
\end{enumerate}
\item The diagrams making up the various matrix elements of interest display different topologies. Indeed, each contribution generated via the application of Wick's theorem can be expressed as a product of {\it strings of contractions}, each of which involves a subset of the $\textbf{L}$ and $\textbf{R}$ operators at play. As for $\langle _s|^i \, _j| ^t \rangle$ with $(i+j) \geq 4$, several such strings actually involve quasi-particle operators belonging to $\textbf{O}^{ij}$, thus forming an overall closed string that is said to be {\it connected} to $\textbf{O}^{ij}$. Translated into diagrammatic language, closed strings correspond to topologically {\it disjoint} closed sub-diagrams. In the case of $\langle _s|^i \, _j| ^t \rangle$, any given diagram is thus made out of disjoint closed sub-diagrams, one of which is connected. As for the matrix elements introduced in Eqs.~\eqref{basiccont}, one set of contractions must form a connected string involving the operators $\beta^{(\dagger)}_{k_2}$ and $\beta^{(\dagger)}_{k_1}$. These two operators cannot belong to two disjoint strings given that any string necessarily involves an even number of quasi-particle operators. Eventually, the diagrams making up the matrix elements introduced in Eqs.~\eqref{basiccont} are made out of disjoint closed sub-diagrams, one of which is connected to the external lines. Last but not least, each diagram contributing to $\langle _s| \, | ^t \rangle$ is made out of disjoint closed sub-diagrams, none of which is connected.
\end{enumerate}

\subsection{Diagrams evaluation}
\label{subs:diag_eval}

Once all the diagrams are drawn, one must compute their expressions. The rules to do so are the following
\begin{enumerate}
\item Label all quasi-particle lines and associate the appropriate factor to each vertex, i.e. a factor $\textbf{z}^*_{l_1l_2}(l)$ ($\textbf{z}^{k_1k_2}(r)$) to each vertex $\textbf{L}$ ($\textbf{R}$) and a factor $\textbf{o}^{k_1 \ldots k_i}_{l_1 \ldots l_j}$ to the vertex $\textbf{O}^{ij}$, respectively.  
\item Sum over all internal line labels. 
\item Include a factor $(n_e!)^{-1}$ for each set of $n_e$ equivalent internal lines. Equivalent internal lines are those connecting identical vertices.
\item For any topologically distinct unlabelled diagram $\Gamma$, a symmetry factor $S_{\Gamma}^{-1}$ must be considered. Given a {\it labelled} version of $\Gamma$, i.e. a version in which each operator $\textbf{L}$ and $\textbf{R}$ carries a specific label, $S_{\Gamma}$ is equal to the number of permutations of the $\textbf{L}$ and $\textbf{R}$ operators delivering a topologically equivalent labelled diagram. The most obvious cases correspond to {\it equivalent subgroups} of $\textbf{L}$ and $\textbf{R}$ operators whose overall permutations lead to topologically equivalent labelled diagrams. The simplest example concerns two $\textbf{L}$ ($\textbf{R}$) operators that are doubly connected to $\textbf{O}^{ij}$ or singly connected to $\textbf{O}^{ij}$ and to the same operator $\textbf{R}$ ($\textbf{L}$). These $n_o\equiv 2$ operators $\textbf{L}$ ($\textbf{R}$) are equivalent and contribute a factor $2!$ to $S_{\Gamma}$. The next simplest example corresponds to $n_o>2$ operators $\textbf{L}$ ($\textbf{R}$) fully connected to $\textbf{O}^{ij}$. These  $n_o>2$ operators are indeed equivalent\footnote{Note that $n_o>2$ operators $\textbf{L}$ ($\textbf{R}$) cannot be equivalent if any of them is connected to an operator $\textbf{R}$ ($\textbf{L}$) given that the latter cannot entertain the same contraction pattern with $n_o>2$ operators $\textbf{L}$ ($\textbf{R}$). More general patterns would however occur if $\textbf{L}$ and $\textbf{R}$ were of higher ranks.} such that their permutations contribute a factor $n_o!$ to $S_{\Gamma}$.  Beyond those two examples, a set of $\textbf{L}$ and $\textbf{R}$ operators can form a string of contractions that is equivalent to other identical strings. Such $n_s$ strings are said to be equivalent and contribute  a factor $n_s!$ to $S_{\Gamma}$. Eventually, less obvious permutations can deliver topologically equivalent labelled diagrams and thus contribute  to $S_{\Gamma}$\footnote{There is no general rule to identify them such that the symmetry factor associated with each topologically distinct unlabelled diagram must be identified on a case by case basis.}.
\item Provide the diagram with a sign $(-1)^{\ell_c}$, where $\ell_c$ is the number of line crossings in the diagram. For diagrams containing external lines, their potential crossing must be counted.
\end{enumerate}

\section{Asymmetric approach}
\label{standardGWT}

The asymmetric approach constitutes the standard path to the off-diagonal Wick theorem at play in the computation of the connected operator kernel~\cite{Balin69wick,RiSc80}. Employing the  simplified notation
\begin{align}
    \textbf{R} &\equiv \textbf{Z}^{20}(l,r) =\frac{1}{2}\sum_{k_1k_2}\textbf{z}^{k_1k_2}(l,r)\beta^\dagger_{k_1}(l)\beta^\dagger_{k_2}(l)\,,
\end{align}
for the Thouless operator introduced in Eq.~\eqref{thoulessoperator3} and satisfying $\langle \Phi(l) | \textbf{R} = 0$, the connected operator kernel reads as
\begin{align}
\frac{\langle \Phi(l) | O | \Phi(r) \rangle}{\langle \Phi(l) | \Phi(r) \rangle} &\equiv \sum_{n=0}^N\sum_{\substack{i,j=0\\i+j=2n}}^{2n} \langle \Phi(l) |\textbf{O}^{ij}  e^{\textbf{R}} |  \Phi(l) \rangle \, ,  \nonumber \\
&= \sum_{n=0}^N\sum_{\substack{i,j=0\\i+j=2n}}^{2n} \langle \Phi(l) | ^{R}{\textbf{O}^{ij}} |  \Phi(l) \rangle \,, \label{SimplifiedsymmetrickernelO}
\end{align}
where the operator
\begin{align}
^{R}{\textbf{O}^{ij}} &\equiv e^{-\textbf{R}} \textbf{O}^{ij} e^{\textbf{R}} \\ 
&= \frac{1}{i!j!}\sum_{\substack{k_1 \ldots k_i\\l_1 \ldots l_j}} \textbf{o}^{k_1\ldots k_i}_{l_1\ldots l_j} \, ^R\beta^{\dagger}_{k_1}  \ldots ^R\beta^{\dagger}_{k_i}   \, ^R\beta_{l_j}  \ldots  ^R\beta_{l_1}   \, , \nonumber
\end{align}
formally reads as $\textbf{O}^{ij}$ but with the quasi-particle operators replaced by their similarity-transformed partners
\begin{align}
    \begin{pmatrix}
      ^R\beta_k\\
      ^R\beta^{\dagger}_k
    \end{pmatrix}
     &\equiv
    e^{-\textbf{R}}\begin{pmatrix}
      \beta_k\\ \beta^\dagger_k
    \end{pmatrix}
    e^{\textbf{R}}
     \, . \label{simtransQPop}
\end{align}

Given that the similarity-transformed quasi-particle operators satisfy anticommutation relations
\begin{subequations}
\label{anticommsimtransop}
\begin{align}
\{^R\beta^{\dagger}_{k_1},^R\beta^{\dagger}_{k_2}\} &= e^{-\textbf{R}}\{\beta^{\dagger}_{k_1},\beta^{\dagger}_{k_2}\}e^{\textbf{R}} = 0\, , \\
\{^R\beta_{k_1},^R\beta_{k_2}\} &= e^{-\textbf{R}}\{\beta_{k_1},\beta_{k_2}\}e^{\textbf{R}} = 0 \, , \\
\{^R\beta_{k_1},^R\beta^{\dagger}_{k_2}\} &= e^{-\textbf{R}}\{\beta_{k_1},\beta^{\dagger}_{k_2}\}e^{\textbf{R}} = \delta_{k_1k_2}\, ,
\end{align}
\end{subequations}
standard Wick's theorem with respect to $| \Phi(l) \rangle$ applies and can be used to compute the matrix elements entering the right-hand side of Eq.~\eqref{SimplifiedsymmetrickernelO}. This results into the standard set of fully contracted terms, except that the elementary contractions at play do not involve the original quasi-particle operators but rather the similarity-transformed ones. The latter are related to the former via a non-unitary Bogoliubov transformation that is now detailed to compute the relevant elementary contractions.

Using Baker-Campbell-Hausdorff (BCH) identity, one first evaluates Eq.~\eqref{simtransQPop} according to 
\begin{align}
^R\beta^{(\dagger)}_{k_1} &= \beta^{(\dagger)}_{k_1} - [\textbf{R},\beta^{(\dagger)}_{k_1}] + \frac{1}{2!} [\textbf{R},[\textbf{R},\beta^{(\dagger)}_{k_1}]] + \ldots\, . \label{BCH}
\end{align}
Given the two elementary commutators
\begin{subequations}
\begin{align}
 \Big[ \beta^{\dagger}_{k}(l) \beta^{\dagger}_{k'}(l) , \beta_{k_1}(l) \Big]   &= \beta^{\dagger}_{k}(l) \delta_{k' k_1} - \beta^{\dagger}_{k'}(l) \delta_{k k_1} \, , \\
 \Big[ \beta^{\dagger}_{k}(l) \beta^{\dagger}_{k'}(l) , \beta^{\dagger}_{k_1} (l)\Big] &= 0 \,,
\end{align}
\end{subequations}
it is straightforward to prove 
\begin{align}
 \Big[ \textbf{R} , \beta_{k_1} \Big]   &= -\sum_{k_2} \Big[U(l) \textbf{z}(l,r)\Big]_{k_1k_2} \beta^{\dagger}_{k_2} (l)\, , \nonumber \\
 \Big[ \textbf{R}, \Big[ \textbf{R} , \beta_{k_1} \Big]  \Big]   &= 0 \, ,  \nonumber  \\
 \Big[ \textbf{R}, \ldots \Big[ \textbf{R} , \beta_{k_1} \Big] \ldots \Big]   &= 0 \, ,  \nonumber \\
 \Big[ \textbf{R} , \beta^{\dagger}_{k_1} \Big] &= -\sum_{k_2} \Big[V(l) \textbf{z}(l,r)\Big]_{k_1k_2} \beta^{\dagger}_{k_2} (l) \, ,  \nonumber \\
 \Big[ \textbf{R}, \Big[ \textbf{R} , \beta^{\dagger}_{k_1} \Big]  \Big]   &= 0 \, ,  \nonumber \\
 \Big[ \textbf{R}, \ldots \Big[ \textbf{R} , \beta^{\dagger}_{k_1} \Big] \ldots \Big]   &= 0 \, ,   \nonumber 
\end{align}
such that
\begin{subequations}
\label{expandedsimtransop}
\begin{align}
^R\beta_{k_1} =&  \sum_{k_2} U_{k_1k_2}(l)  \beta_{k_2} (l)  \nonumber \\
&+ \Big[V^{\ast}(l) + U(l) \textbf{z}(l,r)\Big]_{k_1k_2} \beta^{\dagger}_{k_2} (l) \, , \\
^R\beta^{\dagger}_{k_1} =&   \sum_{k_2} V_{k_1k_2}(l)  \beta_{k_2} (l)  \nonumber \\
&+ \Big[U^{\ast}(l) + V(l) \textbf{z}(l,r)\Big]_{k_1k_2} \beta^{\dagger}_{k_2} (l)  \, ,
\end{align}
\end{subequations}
which can be compacted in matrix form according to the non-unitary Bogoliubov transformation
\begin{align}
    \begin{pmatrix}
      ^R\beta\\
      ^R\beta^{\dagger}
    \end{pmatrix}
     &= 
    \begin{pmatrix}
      U(l) & V^{\ast}(l) + U(l) \textbf{z}(l,r) \\
      V(l) & U^{\ast}(l) + V(l) \textbf{z}(l,r)
    \end{pmatrix}
    \begin{pmatrix}
      \beta(l)\\ \beta^\dagger(l)
    \end{pmatrix}
     \, . \label{simtransQPop2X}
\end{align}

As Eqs.~(\ref{expandedsimtransop}-\ref{simtransQPop2X}) testify, the infinite expansion in Eq.~\eqref{BCH}, originating from the presence of $e^{\textbf{R}}$ in Eq.~\eqref{asymmetricconnectedkernel}, {\it naturally terminates}, i.e. it stops after two terms. Eventually, the four elementary contractions read as
\begin{subequations}
\label{offdiagcontractions1}
\begin{align}
{\rho}_{k_1k_2}(l,r) &= \langle \Phi(l) | ^R{\beta^{\dagger}_{k_2}} \, ^R{\beta_{k_1}} | \Phi(l) \rangle \\
&= \left[V^{\ast}(l)V^T(l) + U(l)\textbf{z}(l,r)V^T(l)\right]_{k_1k_2} \nonumber \\
&=  +{\rho}_{k_1k_2}(l,l)  + \left[U(l)\textbf{z}(l,r)V^T(l)\right]_{k_1k_2} \, ,  \nonumber \\
{\kappa}_{k_1k_2}(l,r) &= \langle \Phi(l) | ^R{\beta_{k_2}} \, ^R{\beta_{k_1}} | \Phi(l) \rangle  \\
&= \left[V^{\ast}(l)U^T(l) + U(l)\textbf{z}(l,r)U^T(l)\right]_{k_1k_2} \nonumber \\
&=  +{\kappa}_{k_1k_2}(l,l) + \left[U(l)\textbf{z}(l,r)U^T(l)\right]_{k_1k_2} \, , \nonumber \\
-\bar{{\kappa}}^{\ast}_{k_1k_2}(l,r) &= \langle \Phi(l) | ^R{\beta^{\dagger}_{k_2}} \, ^R{\beta^{\dagger}_{k_1}} | \Phi(l) \rangle \\
&= \left[U^{\ast}(l)V^T(l) + V(l)\textbf{z}(l,r)V^T(l)\right]_{k_1k_2} \nonumber \\
&=  -\bar{{\kappa}}^{\ast}_{k_1k_2}(l,l) + \left[V(l)\textbf{z}(l,r)V^T(l)\right]_{k_1k_2} \, , \nonumber \\
-{\sigma}^{\ast}_{k_1k_2}(l,r) &= \langle \Phi(l) | ^R{\beta_{k_2}} \, ^R{\beta^{\dagger}_{k_1}} |  \Phi(l) \rangle  \\
&= \left[U^{\ast}(l)U^T(l) + V(l)\textbf{z}(l,r)U^T(l)\right]_{k_1k_2} \nonumber \\
&=  -{\sigma}^{\ast}_{k_1k_2}(l,l) + \left[V(l)\textbf{z}(l,r)U^T(l)\right]_{k_1k_2} \, , \nonumber
\end{align}
\end{subequations}
where Eqs.~(\ref{expandedsimtransop}-\ref{simtransQPop2X}) have been used. This completes the derivation of the off-diagonal Wick theorem where the explicit form of the elementary off-diagonal contractions in Eq.~\eqref{offdiagcontractions1} reflects the asymmetric character of the approach, i.e. the expressions are anchored on the bra state $\langle \Phi(l) |$ and are a functional of the Thouless matrix $\textbf{z}(l,r)$ associated with the transition Bogoliubov transformation of Eqs.~(\ref{transitionBogo3}-\ref{thoulessmatrix3}).

Starting from Eq.~\eqref{offdiagcontractions1} and using repeatedly relations associated with the unitarity of $W(l)$ (Eq.~\eqref{eq:ferm_unitary}), one can symmetrize the elementary contractions by expressing them in terms of the Thouless matrices $\textbf{z}(l)$ and $\textbf{z}(r)$ associated with the left and right states, respectively. Doing so, one recovers exactly Eqs.~\eqref{eq:final1stcontraction} and~\eqref{offdiagelemcontract} obtained directly via the symmetric approach.

\newpage
\begin{strip}

\section{Connected kernel of a rank-3 operator}
\label{examplerank3operator}

According to Eq.~\eqref{eq:gen_kern_tot}, the connected kernel associated with a rank-3 operator $O\equiv \textbf{O}^{[0]}+\textbf{O}^{[2]}+\textbf{O}^{[4]}+\textbf{O}^{[6]}$ reads in terms of the off-diagonal elementary contractions as
\begin{align}
\frac{\langle \Phi(l)| O |\Phi(r) \rangle}{\langle\Phi(l)|\Phi(r) \rangle} &= \textbf{O}^{[0]} + \frac{1}{2} \sum_{k_1k_2} \textbf{o}^{k_1k_2} \bar{{\kappa}}^{\ast}_{k_1k_2}(l,r) + \sum_{k_1l_1} \textbf{o}^{k_1}_{l_1} \rho_{l_1k_1}(l,r)  + \frac{1}{2} \sum_{l_1l_2} \textbf{o}_{l_1l_2} {\kappa}_{l_1l_2}(l,r) \nonumber \\
&+ \frac{1}{8} \sum_{k_1k_2k_3k_4} \textbf{o}^{k_1k_2k_3k_4} \bar{{\kappa}}^{\ast}_{k_1k_2}(l,r) \bar{{\kappa}}^{\ast}_{k_3k_4} (l,r)
+ \frac{1}{2} \sum_{k_1k_2k_3l_1} \textbf{o}^{k_1k_2k_3}_{l_1} \rho_{l_1k_1}(l,r) \bar{{\kappa}}^{\ast}_{k_2k_3}(l,r)   \nonumber \\
&+ \frac{1}{2} \sum_{k_1k_2l_1l_2} \textbf{o}^{k_1k_2}_{l_1l_2}    \rho_{l_1k_1}(l,r)  \rho_{l_2k_2}(l,r) + \frac{1}{4} \sum_{k_1k_2l_1l_2} \textbf{o}^{k_1k_2}_{l_1l_2}   \bar{{\kappa}}^{\ast}_{k_1k_2}(l,r){\kappa}_{l_1l_2} (l,r)\nonumber \\
&+ \frac{1}{2} \sum_{k_1l_1l_2l_3} \textbf{o}^{k_1}_{l_1l_2l_3}   \rho_{l_1k_1} (l,r) {\kappa}_{l_2l_3}(l,r)
+ \frac{1}{8} \sum_{l_1l_2l_3l_4} \textbf{o}_{l_1l_2l_3l_4} {\kappa}_{l_1l_2}(l,r) {\kappa}_{l_3l_4}(l,r)\nonumber \\
&+ \frac{1}{48} \sum_{k_1k_2k_3k_4k_5k_6} \textbf{o}^{k_1k_2k_3k_4k_5k_6} \bar{{\kappa}}^{\ast}_{k_1k_2}(l,r) \bar{{\kappa}}^{\ast}_{k_3k_4} (l,r) \bar{{\kappa}}^{\ast}_{k_5k_6} (l,r) \nonumber \\
&+ \frac{1}{8} \sum_{k_1k_2k_3k_4k_5l_1} \textbf{o}^{k_1k_2k_3k_4k_5}_{l_1}  \rho_{l_1k_1}(l,r) \bar{{\kappa}}^{\ast}_{k_2k_3}(l,r) \bar{{\kappa}}^{\ast}_{k_4k_5}(l,r)    \nonumber \\
&+ \frac{1}{16} \sum_{k_1k_2k_3k_4l_1l_2} \textbf{o}^{k_1k_2k_3k_4}_{l_1l_2} \bar{{\kappa}}^{\ast}_{k_1k_2}(l,r) \bar{{\kappa}}^{\ast}_{k_3k_4}(l,r)    \kappa_{l_1l_2} (l,r)\nonumber \\
&
+ \frac{1}{4} \sum_{k_1k_2k_3k_4l_1l_2} \textbf{o}^{k_1k_2k_3k_4}_{l_1l_2} \rho_{l_1k_1}(l,r) \rho_{l_2k_2}(l,r)  \bar{{\kappa}}^{\ast}_{k_3k_4} (l,r) \nonumber \\
&+ \frac{1}{6} \sum_{k_1k_2k_3l_1l_2l_3} \textbf{o}^{k_1k_2k_3}_{l_1l_2l_3} \rho_{l_1k_1}(l,r) \rho_{l_2k_2}(l,r) \rho_{l_3k_3}(l,r)\nonumber \\
&
+  \frac{1}{4} \sum_{k_1k_2k_3l_1l_2l_3} \textbf{o}^{k_1k_2k_3}_{l_1l_2l_3}  \rho_{l_1k_1}(l,r) \bar{{\kappa}}^{\ast}_{k_2k_3}(l,r) \kappa_{l_2l_3}(l,r)  \nonumber \\
&+ \frac{1}{16} \sum_{k_1k_2l_1l_2l_3l_4} \textbf{o}^{k_1k_2}_{l_1l_2l_3l_4} \bar{{\kappa}}^{\ast}_{k_1k_2}(l,r)  \kappa_{l_1l_2} (l,r) \kappa_{l_3l_4} (l,r)\nonumber \\
&
+ \frac{1}{4} \sum_{k_1k_2l_1l_2l_3l_4} \textbf{o}^{k_1k_2}_{l_1l_2l_3l_4} \rho_{l_1k_1}(l,r) \rho_{l_2k_2}(l,r)  \kappa_{l_3l_4}(l,r)  \nonumber \\
&+ \frac{1}{8} \sum_{k_1l_1l_2l_3l_4l_5} \textbf{o}^{k_1}_{l_1l_2l_3l_4l_5} \rho_{l_1k_1}(l,r)  \kappa_{l_2l_3}(l,r)  \kappa_{l_4l_5}(l,r)  \nonumber \\
&+\frac{1}{48} \sum_{l_1l_2l_3l_4l_5l_6} \textbf{o}_{l_1l_2l_3l_4l_5l_6}  \kappa_{l_1l_2}(l,r)  \kappa_{l_3l_4}(l,r)  \kappa_{l_5l_6}(l,r) \, . \label{connectedopkernelrank3}
\end{align}
\end{strip}

\end{appendix}

\bibliography{biblio.bib}

\end{document}